\documentclass[aps,prd,nofootinbib,twocolumn,superscriptaddress,letterpaper,preprintnumbers]{revtex4-1}
\usepackage{amsmath, amssymb,braket, mathtools, soul}
\usepackage[dvipsnames]{xcolor}
\usepackage[svgnames]{xcolor}
\colorlet{energygreen}{green!60!black}
\definecolor{MathematicaBlue}{RGB}{94,129,181}
\definecolor{MathematicaOrange}{RGB}{255,140,0}
\usepackage{color}
\usepackage{float}
\usepackage{graphicx}
\usepackage[caption=false]{subfig}
\usepackage{natbib}
\usepackage{comment}
\usepackage[colorlinks=true
,urlcolor=blue
,anchorcolor=blue
,citecolor=blue
,filecolor=blue
,linkcolor=red
,menucolor=blue
,hyperfootnotes=false,
,linktocpage=true
,pdfproducer=medialab
,pdfa=true
]{hyperref}

\usepackage{appendix}

\usepackage{tikz}
\usetikzlibrary{calc,decorations.pathmorphing,arrows.meta}

\tikzset{
  infl/.style={
    black,
    line cap=round,
    line width=.7pt
  },
  boundary/.style={
    gray!60,
    line cap=round,
    line width=2.0pt
  },
  vertex/.style={
    circle,
    fill=black,
    inner sep=1.1pt
  },
  source/.style={
    black,
    line width=.7pt,
    dash pattern=on 4pt off 3pt,
    line cap=round
  },
  sourceX/.style={
    black,
    line width=.85pt,
    line cap=round
  },
  energycut/.style={
    green!60!black,
    line width=2.2pt,
    line cap=round
  },
  gravitonwave/.style={
    black,
    line width=.7pt,
    decorate,
    decoration={snake, amplitude=1.15pt, segment length=6.5pt}
  }
}

\newcommand{\MomentumTriangleVar}[3][1]{%
\begin{tikzpicture}[
  scale=#1,
  transform shape,
  line cap=round,
  line join=round,
  every node/.style={font=\small}
]
  \pgfmathsetmacro{\L}{3.2}

  \pgfmathsetmacro{\xthree}{#2}          
  \pgfmathsetmacro{\thetarad}{#3}        
  \pgfmathsetmacro{\thetadeg}{deg(\thetarad)} 

  \pgfmathsetmacro{\alpha}{180 - \thetadeg}

  \pgfmathsetmacro{\xtwo}{sqrt(1 + \xthree*\xthree + 2*\xthree*cos(\thetadeg))}

  \coordinate (O) at (0,0);
  \coordinate (A) at (\L,0);
  \coordinate (B) at ({\L*\xthree*cos(\alpha)},{\L*\xthree*sin(\alpha)});

  \pgfmathsetmacro{\rangle}{min(0.80,max(0.22,0.34*\L*\xthree))}

  \draw[black, very thick]
    (O) -- (A)
    node[midway, below=3pt] {$k_1$};

  \draw[black, very thick, dashed]
    (O) -- (B)
    node[midway, above left=2pt] {$k_3=#2\,k_1$};

  \draw[black, very thick, dashed]
    (A) -- (B)
    node[pos=0.56, above right=2pt]
    {$k_2 \simeq \pgfmathprintnumber[fixed,precision=2]{\xtwo}\,k_1$};

  \fill (O) circle (1.5pt);
  \fill (A) circle (1.5pt);
  \fill (B) circle (1.5pt);

  \draw[black, thin]
    (\rangle,0) arc[start angle=0,end angle=\alpha,radius=\rangle];

  \node at ({0.85*\rangle*cos(\alpha/2)},{0.85*\rangle*sin(\alpha/2)})
    {};

\end{tikzpicture}%
}

\newcommand{\DoubleWavyLine}[2]{%
  \draw[gravitonwave]
    ($(#1)!1.25pt! 90:(#2)$) -- ($(#2)!1.25pt!-90:(#1)$);
  \draw[gravitonwave]
    ($(#1)!1.25pt!-90:(#2)$) -- ($(#2)!1.25pt! 90:(#1)$);
}

\newcommand{\EnergyInjectionBispectrum}[1][1]{%
\begin{tikzpicture}[
  scale=#1,
  transform shape,
  line cap=round,
  line join=round,
  baseline={(current bounding box.center)}
]
  \def\yb{2.0}

  \coordinate (VL) at (-1.25,0.5);
  \coordinate (VR) at ( 1.45,0.5);

  \coordinate (Kone)   at (-2.55,\yb);
  \coordinate (Ktwo)   at (-1.25,\yb);
  \coordinate (Kthree) at ( 2.05,\yb);

  \coordinate (Xsrc) at ($(VR)+(45:1.35)$);

  \draw[infl] (VL) -- (Kone);
  \draw[infl] (VL) -- (Ktwo);
  \draw[infl] (VR) -- (Kthree);

  \DoubleWavyLine{VL}{VR}

  \draw[source] (VR) -- (Xsrc);

  \draw[sourceX] ($(Xsrc)+(-0.11, 0.11)$) -- ($(Xsrc)+( 0.11,-0.11)$);
  \draw[sourceX] ($(Xsrc)+(-0.11,-0.11)$) -- ($(Xsrc)+( 0.11, 0.11)$);

  \node[vertex] at (VL) {};
  \node[vertex] at (VR) {};

  \node[font=\large, anchor=south] at ($(Kone)+(0,0.0)$) {$\delta\sigma(\mathbf{k}_1)$};
  \node[font=\large, anchor=south] at ($(Ktwo)+(0,0.0)$) {$\delta\sigma(\mathbf{k}_2)$};
  \node[font=\large, anchor=south] at ($(Kthree)+(0,0.0)$) {$\delta\sigma(\mathbf{k}_3)$};

  \node[
    font=\large,
    fill=white,
    inner sep=1.0pt
  ] at ($(VL)!0.5!(VR)+(0,0.48)$) {$\widetilde h_{\mu\nu},\,\varphi$};

  \node[font=\large, anchor=west] at ($(VR)!0.58!(Xsrc)+(0.08,0.0)$) {$\dot{\sigma}_0$};

  \draw[boundary] (-3.25,\yb) -- (3.35,\yb);

\end{tikzpicture}%
}

\newcommand{\EnergyInjectionBispectrumMM}[1][1]{%
\begin{tikzpicture}[
  scale=#1,
  transform shape,
  line cap=round,
  line join=round,
  baseline=(etaBase.base)
]
  \def\yb{2.0}

  \node[inner sep=0pt] (etaBase) at (0,\yb) {};

  \coordinate (VL) at (-1.25,3.15);
  \coordinate (VR) at ( 1.45,3.15);

  \coordinate (Kone)   at (-2.55,\yb);
  \coordinate (Ktwo)   at (-1.25,\yb);
  \coordinate (Kthree) at ( 2.05,\yb);

  \coordinate (Xsrc) at ($(VR)+(-45:1.15)$);

  \draw[infl] (VL) -- (Kone);
  \draw[infl] (VL) -- (Ktwo);
  \draw[infl] (VR) -- (Kthree);

  \DoubleWavyLine{VL}{VR}

  \draw[source] (VR) -- (Xsrc);

  \draw[sourceX] ($(Xsrc)+(-0.11, 0.11)$) -- ($(Xsrc)+( 0.11,-0.11)$);
  \draw[sourceX] ($(Xsrc)+(-0.11,-0.11)$) -- ($(Xsrc)+( 0.11, 0.11)$);

  \node[vertex] at (VL) {};
  \node[vertex] at (VR) {};

  \draw[boundary] (-3.25,\yb) -- (3.35,\yb);

\end{tikzpicture}%
}

\newcommand{\EnergyInjectionBispectrumPM}[1][1]{%
\begin{tikzpicture}[
  scale=#1,
  transform shape,
  line cap=round,
  line join=round,
  baseline={(current bounding box.center)}
]
  \def\yb{2.0}

  \coordinate (VL) at (-1.25,0.50);
  \coordinate (VR) at ( 1.45,3.05);

  \coordinate (Kone)   at (-2.55,\yb);
  \coordinate (Ktwo)   at (-1.25,\yb);
  \coordinate (Kthree) at ( 2.05,\yb);

  \coordinate (Xsrc) at ($(VR)+(-45:1.15)$);

  \draw[infl] (VL) -- (Kone);
  \draw[infl] (VL) -- (Ktwo);
  \draw[infl] (VR) -- (Kthree);

  \DoubleWavyLine{VL}{VR}

  \draw[source] (VR) -- (Xsrc);

  \draw[sourceX] ($(Xsrc)+(-0.11, 0.11)$) -- ($(Xsrc)+( 0.11,-0.11)$);
  \draw[sourceX] ($(Xsrc)+(-0.11,-0.11)$) -- ($(Xsrc)+( 0.11, 0.11)$);

  \node[vertex] at (VL) {};
  \node[vertex] at (VR) {};

  \draw[boundary] (-3.25,\yb) -- (3.35,\yb);

\end{tikzpicture}%
}

\newcommand{\EnergyInjectionBispectrumMP}[1][1]{%
\begin{tikzpicture}[
  scale=#1,
  transform shape,
  line cap=round,
  line join=round,
  baseline={(current bounding box.center)}
]
  \def\yb{2.0}

  \coordinate (VL) at (-1.25,3.05);
  \coordinate (VR) at ( 1.45,0.50);

  \coordinate (Kone)   at (-2.55,\yb);
  \coordinate (Ktwo)   at (-1.25,\yb);
  \coordinate (Kthree) at ( 2.05,\yb);

  \coordinate (Xsrc) at ($(VR)+(45:1.15)$);

  \draw[infl] (VL) -- (Kone);
  \draw[infl] (VL) -- (Ktwo);
  \draw[infl] (VR) -- (Kthree);

  \DoubleWavyLine{VL}{VR}

  \draw[source] (VR) -- (Xsrc);

  \draw[sourceX] ($(Xsrc)+(-0.11, 0.11)$) -- ($(Xsrc)+( 0.11,-0.11)$);
  \draw[sourceX] ($(Xsrc)+(-0.11,-0.11)$) -- ($(Xsrc)+( 0.11, 0.11)$);

  \node[vertex] at (VL) {};
  \node[vertex] at (VR) {};

  \draw[boundary] (-3.25,\yb) -- (3.35,\yb);

\end{tikzpicture}%
}

\newcommand{\mpl}{M_{\rm pl}}

\newcommand{\D}{{\rm d}}

\newcommand{\fnl}{f_{\rm NL}}

\let\vec\mathbf
\newcommand{\es}[2] {\begin{equation} \label{#1} \begin{split} #2 \end{split} \end{equation}}

\definecolor{blue3}{RGB}{31,119,180}
\definecolor{green3}{RGB}{44,160,44}

\begin{document}
\title{First Search for Kaluza-Klein Gravitons and Radion Using {\it Planck} Data}
\author{Alexander P. Cassem} 
\affiliation{Institute of Cosmology, Department of Physics and Astronomy, Tufts University, Medford, MA 02155, USA}
\author{Soubhik Kumar} 
\affiliation{Institute of Cosmology, Department of Physics and Astronomy, Tufts University, Medford, MA 02155, USA}
\begin{abstract}
Heavy moduli and Kaluza-Klein (KK) gravitons from extra dimensions may evade terrestrial probes but can be produced during inflation, generating primordial non-Gaussianity (NG) through unavoidable couplings to density perturbations. 
In a warped five-dimensional (5D) model, we compute the full radion- and KK-graviton-mediated bispectra and perform the first search for these signals using {\it Planck} 2018 temperature and polarization data.
We find no significant evidence for NG, with the maximum significance being $1.8\sigma$ for $m_{\rm KK}\approx 1.6H$.
We also identify a 5D setup which naturally generates NG with $\fnl\sim 1-50$, within the reach of future surveys.
\end{abstract}
\maketitle
\section{Introduction and Summary}\label{sec: introduction and summary}
Extra spatial dimensions often arise in efforts to go beyond the Standard Model (SM) of particle physics.
In string theory, extra dimensions are required to have a mathematically consistent description of quantum gravity~\cite{Green:1987sp, Polchinski:1998rq}.
Extra dimensions could also play central roles in explaining a number of key puzzles beyond the SM, such as the origin of the Higgs mass~\cite{Arkani-Hamed:1998jmv, Antoniadis:1998ig, Randall:1999ee, Randall:1999vf}, gauge coupling unification~\cite{Dienes:1998vg, Kawamura:2000ev, Hall:2001pg}, the origin of flavor hierarchies and neutrino masses~\cite{Grossman:1999ra, Gherghetta:2000qt}, and the identity of dark matter~\cite{Servant:2002aq, Cheng:2002ej}.
Accordingly, extensive searches for signatures of extra dimensions have been performed in astrophysics, cosmology, and especially particle colliders~\cite{ParticleDataGroup:2024cfk}.
In general, these searches show that the scale of higher-dimensional quantum gravity (for flat extra dimensions~\cite{Arkani-Hamed:1998jmv, Antoniadis:1998ig}) or  Kaluza-Klein (KK) masses (for warped extra dimensions~\cite{Randall:1999ee, Randall:1999vf}) have to be above ${\cal O}(1-100)$~TeV, depending on the specific set up.
However, these searches lose their sensitivity if the extra dimensions are much smaller, or equivalently the KK masses are much larger, simply because the associated energy scales cannot be reached in terrestrial experiments.

The early Universe provides a natural way around this challenge since the energy scales then could have been much higher than the TeV scale.
This is especially true during cosmic inflation when the Hubble scale $H$, that governs the spacetime expansion rate, could have been as high as $4\times 10^{13}$~GeV~\cite{BICEP:2021xfz}.
Due to the rapidly expanding spacetime, states with masses $\sim H$ can be produced via vacuum fluctuations, without needing any special coupling in the theory.
Therefore, if some of the KK gravitons or moduli---always present in extra-dimensional theories---have masses $\sim H$, they could already have been produced during inflation, despite being too heavy to be accessed in terrestrial experiments today.

Importantly, since KK gravitons and moduli are gravitational degrees of freedom, they have irreducible couplings to the source responsible for generating the scalar density perturbations $\zeta$ observed in the cosmic microwave background (CMB) and large-scale structure (LSS).
Thus, following their production during inflation, KK gravitons and moduli can decay into $\zeta$, leading to primordial non-Gaussianity (NG).
The simplest such NG is the bispectrum $\langle \zeta(\vec{k}_1) \zeta(\vec{k}_2) \zeta(\vec{k}_3)\rangle$ as a function of (comoving) three-momenta $\vec{k}_{1,2,3}$.
The momentum dependence of such bispectra encodes both the mass and the spin of the heavy particle.
For example, a KK graviton-mediated bispectrum contains oscillations, with a frequency determined by the KK graviton mass, and a distinctive angular dependence that is characteristic of its spin-2 nature.
Thus, searches for such NG let us do mass-spin spectroscopy at very high energies, potentially at $10^{13}-10^{14}$~GeV, with the early Universe functioning as a `cosmological collider'~\cite{Chen:2009zp, Baumann:2011su, Chen:2012ge, Noumi:2012vr, Arkani-Hamed:2015bza}.
This provides a direct probe of ultraviolet (UV) physics far beyond the reach of terrestrial experiments.

This opportunity has motivated a number of recent searches for cosmological collider signals, from both scalar and spinning massive particles, using CMB and LSS data~\cite{Green:2023uyz, Cabass:2024wob, Goldstein:2024bky, Sohn:2024xzd, Philcox:2025wts, Bao:2025onc, Philcox:2025bbo, Suman:2025vuf, Suman:2025tpv, Green:2026yev, Kumar:2026ogn, Philcox:2026njr, Kumar:2026dih}.
In particular, the existing searches for spin-2 particles have considered only dimension-6 effective field theory (EFT) operators that describe the (cubic) interaction between $\zeta$ and spin-2 particles. 
However, KK gravitons generate the leading NG already at dimension-5. 
Within EFT power counting, these dimension-5 signals are parametrically larger than the dimension-6 contributions previously considered.
Additionally, the dimension-5 and dimension-6 operators generate distinct bispectrum shapes.
Although these dimension-5 operators for KK gravitons were identified in~\cite{Kumar:2018jxz, Kumar:2025anx}, their full bispectrum shapes have not previously been computed, and existing cosmological collider searches therefore miss the leading NG signatures of KK gravitons.
Here we compute these shapes and perform the first dedicated search using {\it Planck} 2018 temperature and polarization data.

\begin{figure}
\centering
    \includegraphics[width=0.95\linewidth]{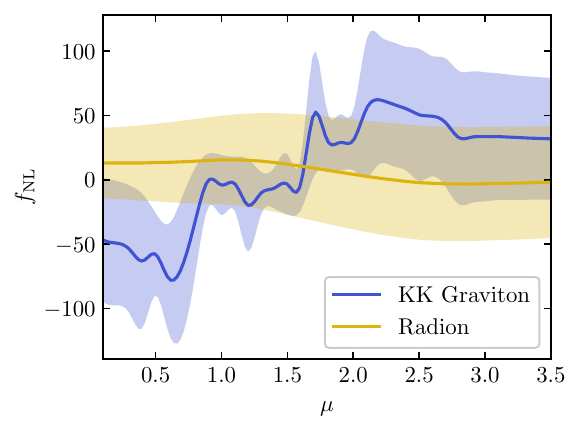}
    \caption{Estimated strength of the bispectrum ($\fnl$), derived in this work using the {\it Planck} 2018 data, due to radion and KK graviton. The parameter $\mu=\sqrt{M^2/H^2-9/4}$ is determined by the heavy particle mass $M$ and the inflationary Hubble scale $H$. These results are consistent with $\fnl=0$ at 95\% CL, with the error bands at 68\% CL.}
    \label{fig: fNL vs mu plot}
\end{figure}

The results of our search for the radion modulus and the KK graviton are shown in Fig.~\ref{fig: fNL vs mu plot}, where we have parametrized the overall strength of the bispectrum using the $\fnl$ parameter, defined in Eq.~\eqref{eq:fnl}.
We do not find any evidence for these shapes at 95\% CL.
To interpret these values of $\fnl$, as well as derive an explicit target for future searches, particularly with LSS data, we outline a warped extra-dimensional construction where the spacetime is a slice of five-dimensional anti-de Sitter space (${\rm AdS}_5$), with UV and infrared (IR) boundaries~\cite{Randall:1999ee}.
In this setup, the wave functions of the radion and KK gravitons peak towards the IR boundary.
Thus, if density perturbations $\zeta$ originate from a field $\sigma$ that is also localized on the IR boundary, the overlap of the corresponding wave functions naturally enhances the resulting NG.
As a concrete realization of this possibility, we focus on the scenario of modulated reheating~\cite{Dvali:2003em} where $\sigma$ modulates the reheating surface and determines $\zeta$.
This scenario naturally allows $\fnl^{\rm rad,~KK} \sim 1-50$, while remaining within the regime of theoretical control.
Such values of $\fnl$ are consistent with our {\it Planck} bounds and could be probed more effectively with future LSS observations.

\section{Four Dimensional Effective Theory}\label{sec: four dimensional effective theory}
To keep our discussion model independent, we start with an effective 4D description, postponing the detailed 5D construction.
This will be sufficient for computing the {\it shapes} of the bispectrum, which are determined by the structure of the operators coupling the KK gravitons and the radion to $\sigma$.
These structures are entirely fixed by symmetries, as we discuss shortly.
On the other hand, computing the magnitude of the bispectra requires a knowledge of the actual KK graviton and radion couplings, which we will determine from the 5D setup.

Our 4D description follows from the $\text{AdS}/\text{CFT}$ duality~\cite{Maldacena:1997re, Gubser:1998bc, Witten:1998qj} which relates a 5D theory in $\text{AdS}_5$ in terms of a large-$N$, 4D strongly coupled conformal field theory ($\text{CFT}_4$)~\cite{Arkani-Hamed:2000ijo, Rattazzi:2000hs}.
This 4D theory is characterized by two dimensionful scales, the Planck scale $\mpl$ and a confinement scale $\Lambda_c$.
The interactions among the composite states, dual to IR-localized 5D fields, are controlled by $N$ and $\Lambda_c$.
Cosmology introduces two additional scales: the inflationary Hubble scale $H$ and the mass of $\sigma$, $m_\sigma$.

We are interested in describing the confining phase of this theory during inflation and thus we require $H<\Lambda_c\lesssim m_{\rm KK}$, where the last relation follows from the duality between the KK modes and the composite states.
An additional restriction comes from the fact that the distinctive oscillatory NG mediated by KK gravitons become exponentially suppressed if their masses $m_{\rm KK} \gg H$, owing to the fact that vacuum fluctuations cannot efficiently produce {\it on-shell} KK gravitons for $m_{\rm KK} \gg H$. 
Therefore, in this work we focus on scenarios where the lightest KK gravitons have $m_{\rm KK}\simeq \Lambda_c \simeq H$.
However, this is not a fundamental restriction since there are mechanisms with sources of energy injection, such as a chemical potential~\cite{Bodas:2020yho, Hook:2019vcn, You:2026xoq, Wang:2020ioa, Bodas:2024hih} or a primordial feature~\cite{Chen:2022vzh}, that can produce on-shell particles parametrically heavier than $H$.

Our focus on $\Lambda_c \simeq H$ also implies that the inflationary energy density $V_{\rm inf}\sim H^2\mpl^2 \gg \Lambda_c^4$.
This indicates that the field(s) driving the homogeneous inflationary expansion cannot be composite.
However, for the composite states (dual to radion and KK graviton) to mediate observable NG, density perturbations $\zeta$ should also be sourced by some composite state, which we label as $\sigma$.
This feature of one (elementary) source driving the homogeneous inflationary expansion and another (composite) source giving rise to $\zeta$ is naturally present in several mechanisms, e.g., involving curvaton~\cite{Lyth:2001nq, Enqvist:2001zp, Moroi:2001ct}, modulated reheating~\cite{Dvali:2003em, Zaldarriaga:2003my}, and inhomogeneous end of inflation~\cite{Bernardeau:2002jf, Bernardeau:2004zz}.
For concreteness, we will focus on the modulated reheating scenario to derive $\fnl$.

Since $\zeta$ is controlled by $\sigma$, its mass $m_\sigma$ should be smaller than $H$, and hence $\Lambda_c$, such that it can obtain (approximately) scale-invariant, long-wavelength fluctuations.
This can happen naturally if $\sigma$ arises as a composite pseudo-Nambu Goldstone boson, similar to pions in the chiral Lagrangian, and respects a shift symmetry, $\sigma\rightarrow \sigma+{\rm constant}$, that is softly broken only by $m_\sigma$.

With these ingredients and symmetries, the interaction Lagrangian relevant for NG is
\es{eq:int}{
{{\cal L}_{\rm int}\over \sqrt{-\bar{g}}} = -{1\over 2}\left( {c_g \over \Lambda_c} \tilde{h}_{\mu\nu}\nabla^\mu\sigma \nabla^\nu\sigma + {c_r \over \Lambda_c} \delta\varphi \nabla^\mu\sigma \nabla_\mu\sigma\right),
}
which will also be derived in Eq.~\eqref{eq:int5D} using 5D. Here the background metric is approximately de Sitter (dS) $\D s^2_4 = \bar{g}_{\mu\nu}\D x^\mu \D x^\nu =  -\D t^2 + e^{2Ht} \D \vec{x}^2 = (-\D \eta^2 + \D \vec{x}^2)/(H^2 \eta^2)$, where $t$ and $\eta$ denote physical and conformal times, respectively.
Here, $\tilde{h}_{\mu\nu}$ is a spin-2 composite resonance, dual to a KK graviton, and the above includes the leading interactions with $\sigma$.
The dilaton $\varphi$ (with fluctuations $\delta\varphi$), dual to the radion modulus, couples to the 4D stress-energy tensor (see, e.g.,~\cite{Goldberger:2007zk}), giving the interaction term in Eq.~\eqref{eq:int}.
The coefficients $c_g$ and $c_r$ are proportional to $1/N$ in the 't Hooft limit~\cite{Witten:1979kh}.
We have not kept terms not obeying $\sigma$ shift symmetry, since the associated coefficients would be suppressed by $m_\sigma \ll H$ and contribute to subleading NG signals.

To compute the bispectra using Eq.~\eqref{eq:int}, we split the field $\sigma(t, \vec{x})$ in terms of a homogeneous component $\sigma_0(t)$ and its fluctuations $\delta\sigma(t, \vec{x})$, as $\sigma = \sigma_0 + \delta\sigma$.
Eq.~\eqref{eq:int} then gives both quadratic mixing and cubic interaction terms,
\begin{align}
    &\mathcal{L}_{\rm mix} ={c_g\over \Lambda_c}  {\tilde{h}_{\eta\eta} \over H^2 \eta^2} (\delta\sigma \ddot{\sigma}_0-\delta\sigma\dot{\sigma}_0 H) - {c_r\over \Lambda_c}  {\delta\varphi \over H^4 \eta^4} \dot{\delta\sigma}\dot{\sigma}_0,\label{eq: quadratic mixing int}\\
    &\mathcal{L}_{\rm cubic} = 
    - \frac{c_g}{2\Lambda_c}\, \left[\tilde{h}_{\eta\eta}   (\partial_\eta \delta\sigma)^2 - 2\tilde{h}_{i\eta}   \partial_i \delta\sigma \partial_\eta \delta\sigma\right.\nonumber \\ &\left.+ \tilde{h}_{ij}   \partial_i \delta\sigma \partial_j \delta\sigma\right] - \frac{c_r}{2\Lambda_c}{\delta\varphi \over H^2 \eta^2}\left[(\partial_i \delta\sigma)^2-(\partial_\eta \delta\sigma)^2 \right].\label{eq: cubic mixing}
\end{align}
Here an overdot denotes a derivative with respect to $t$.
The mixing between the temporal component $\tilde{h}_{\eta\eta}$ and $\delta\sigma$ involves both $\ddot{\sigma}_0$ and $H\dot{\sigma}_0$.
However, because $m_\sigma\ll H$, $\sigma_0$ undergoes slow-roll and $|\ddot{\sigma}_0| \ll H|\dot{\sigma}_0|$.
Therefore, we only consider the contribution proportional to $\dot{\sigma}_0$.
The terms in $\mathcal{L}_{\rm mix}$ and $\mathcal{L}_{\rm cubic}$ induce a three-point correlation function $\langle\delta\sigma(\vec{k}_1) \delta\sigma(\vec{k}_2) \delta\sigma(\vec{k}_3)\rangle \equiv (2\pi)^3 \delta^{3}(\vec{k}_1+\vec{k}_2+\vec{k}_3) \langle\delta\sigma(\vec{k}_1) \delta\sigma(\vec{k}_2) \delta\sigma(\vec{k}_3)\rangle'$ that we compute using the `in-in' formalism~\cite{Weinberg:2005vy}.
The associated in-in diagram, for the `s-channel' permutation, is the following
\begin{equation}
    \EnergyInjectionBispectrum[0.8]\nonumber
\end{equation}
where one external leg is set to $\dot{\sigma}_0$. Here, the exchange particle (wavy line) could either be the dilaton $\varphi$ or the helicity-0 component of the spin-2 resonance $\tilde{h}_{\mu\nu}$.
Along with the above diagram, all other permutations should also be added to compute the full bispectrum; we provide the details in Appendix~\ref{subsec: in-in propagators}. 
We note that the KK-graviton interactions from the operator in Eq.~\eqref{eq:int} are not present in the previous cosmological collider studies of massive spin-2 particles, although these can be derived using the EFT of inflation~\cite{Cheung:2007st} approach.
We discuss this further in Appendix~\ref{app: EFTofInf argument for spin-s}.
For additional studies of spin-2, see, e.g., Refs.~\cite{Tong:2022cdz, Hubisz:2024xnj, Bao:2025onc}.

\section{Shape Function and Search with {\it Planck} Data}\label{sec: shape function}
To search for NG in the {\it Planck} data, we treat the conversion $\delta\sigma\to\zeta$ through an overall linear factor, $\zeta(\vec{k}) = f \delta\sigma(\vec{k})$, where $f$ is model dependent.
Since $f$ does not affect the kinematic dependence of the three-point function, we leave an explicit expression for $f$ for later.
The three-point function of $\delta\sigma$, $\langle\delta\sigma(\vec{k}_1) \delta\sigma(\vec{k}_2) \delta\sigma(\vec{k}_3)\rangle' \equiv B(k_1, k_2, k_3)$, thus induces a bispectrum of $\zeta$
\es{eq:fnl}{
\left\langle \zeta(\vec{k}_1)\zeta(\vec{k}_2)\zeta(\vec{k}_3) \right\rangle' = \frac{72}{5}\,\pi^4 A_s^2 f_{\mathrm{NL}} 
\frac{S(k_1,k_2,k_3)}{k_1^2 k_2^2 k_3^2},
}
where $A_s = (k^3/(2\pi^2)) (k/k_*)^{1-n_s}\langle\zeta(\vec{k})\zeta(-\vec{k})\rangle' \approx 2.1\times 10^{-9}$~\cite{SPT-3G:2025bzu} is the dimensionless power spectrum.
The kinematic dependence of the bispectrum is encoded in the shape function $S$, while its strength is given by $\fnl$.
A normalization of $S$ should be fixed to uniquely specify $\fnl$.
In this work, we follow the convention in~\cite{Kumar:2026dih, Kumar:2026ogn}, and fix the normalization by setting the maximum of $S$, in the entire physical kinematic space, to be unity.

\begin{figure}
    \centering
    \includegraphics[width=0.9\linewidth]{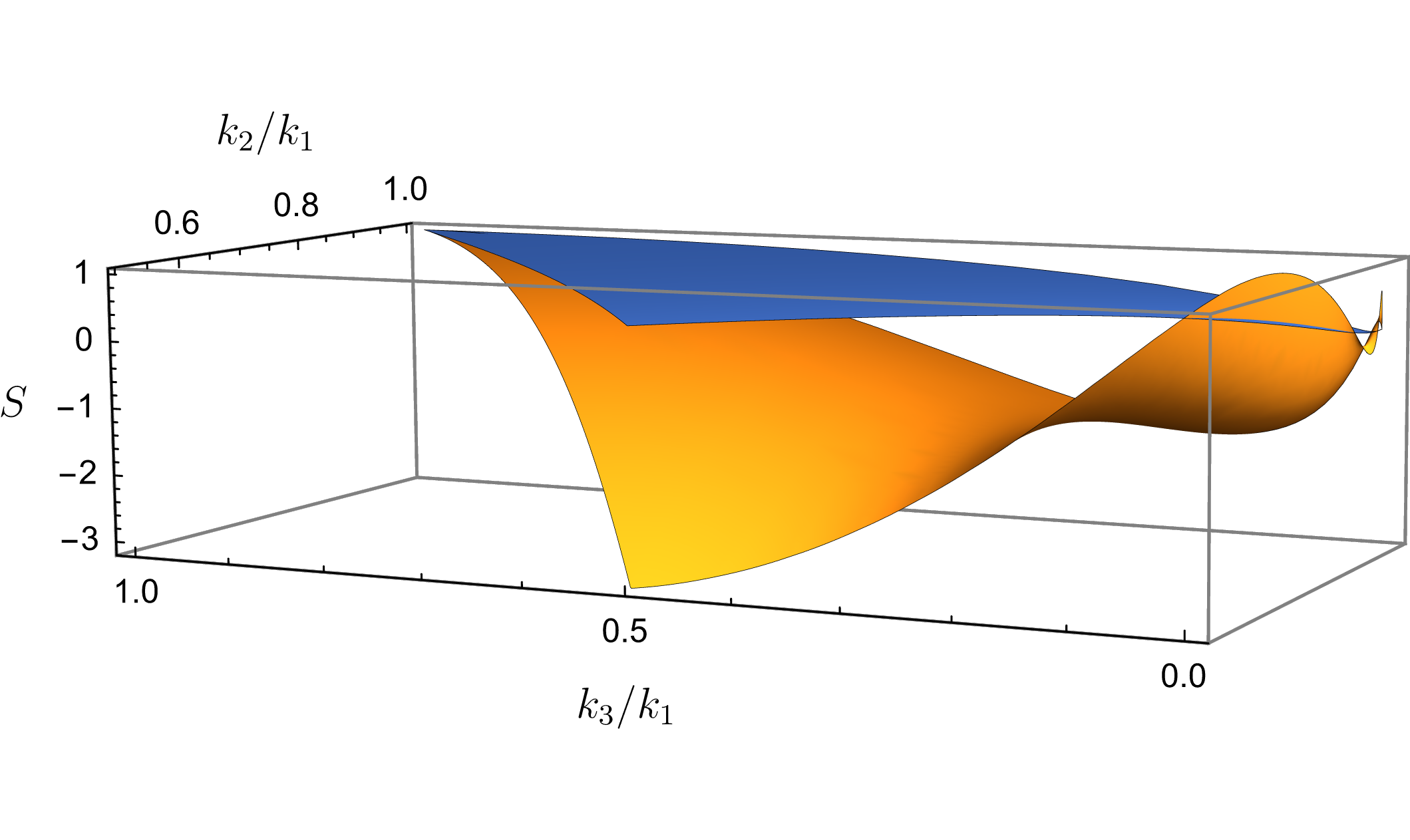}
    \caption{Shape functions for the KK graviton in {\color{orange}{orange}} with $\mu = 1.1$ and the radion in {\color{blue}{blue}} with $\mu = 1.1$, normalized such that $S=1$ at the equilateral point $k_1=k_2=k_3=1$.
    }
    \label{fig:shape}
\end{figure}

We use the {\tt CMB-BEST} package~\cite{Sohn:2023fte} to search for NG in the {\it Planck} 2018 temperature and polarization data.
We input our numerical shape function ${\cal S}_{\rm in} = \fnl S = \fnl k_1^2 k_2^2 k_3^2 B(k_1, k_2, k_3)/{\rm Max} [|k_1^2 k_2^2 k_3^2 B(k_1, k_2, k_3)|]$ to {\tt CMB-BEST}, use the {\tt custom\_shape\_evals} option, and retrieve the estimated $\fnl$, along with the uncertainties.
We set $p_{\rm max}$, the maximum order of Legendre polynomials in the basis expansion, to 30.
To carry out the above steps, we need to know $S$ in the whole kinematic space, and we perform a numerical evaluation of the `in-in' diagrams to obtain that (Appendix~\ref{app sec: Details on Mode Functions}).
A direct computation in the full 3D space is computationally expensive.
However, since $S$ is scale invariant, it depends only on two momentum ratios.
Accordingly, we first evaluate $S$ in a grid of $120\times 120$ points, $\log_{10}$-uniformly spaced in the $(k_2/k_1, k_3/k_1)$ plane.
Subsequently, we use the scale invariant nature of our shape to obtain $S$ in the 3D $(k_1, k_2, k_3)$ space, satisfying $10^{-3}\leq k_{1,2,3}/k_{\rm max}\leq 1$, with $k_{\rm max} = 0.209/{\rm Mpc}$, the dynamical range of {\tt CMB-BEST}.
Finally, we evaluate $S$ for 35 equally spaced points for $0.1 \leq \mu \leq 3.5$, for both radion and KK graviton-mediated shapes.
While computing $S$ for KK gravitons become increasingly time consuming with larger $\mu$, we have checked that our bounds on $\fnl$ become insensitive to $\mu$ for $\mu\gtrsim 3$.
In Fig.~\ref{fig:shape}, we show example shape functions for the radion and KK graviton-mediated NG.

Our results for $\fnl$ are shown in Fig.~\ref{fig: fNL vs mu plot}.
For all the considered values of the radion mass, $\fnl$ is consistent with zero at 68\% CL.
For KK graviton, however, we find $\fnl\neq 0$ at 68\% CL, for a range of masses.
In particular, the maximum local significance ($1.8\sigma$) for $\fnl\neq 0$ is reached at $\mu=0.5$, for which $\fnl = -58\pm 32$ at 68\% CL.
For $\mu\gtrsim 2.6$, the bounds flatten, which is expected since for these values of $\mu$, on-shell production of the radion or the KK graviton is exponentially suppressed as $\exp(-\pi \mu)$.
Thus the leading NG contribution comes from a contact diagram where the massive particle has been integrated out.

\section{ Five Dimensional Geometry}\label{sec: five dimensional geometry and model interpretation}
The bounds $|\fnl|\lesssim 100$ in Fig.~\ref{fig: fNL vs mu plot} are valid as long as the radion (dilaton) and the KK graviton (spin-2 resonance) have a coupling to $\sigma$, and hence $\zeta$, as in Eq.~\eqref{eq:int}.
In particular, these observational bounds are independent of the specific values $c_g, c_r, \Lambda_c$.
However, to assess whether $\fnl\sim 1-50$ can realistically arise and be probed in future searches, we now consider an explicit 5D construction to compute $c_g, c_r, \Lambda_c$ and $\fnl$.

We consider a Randall-Sundrum-1 (RS1) geometry~\cite{Randall:1999ee} with an ${\rm AdS}_5$ spacetime where the extra dimension is an interval $y \in (-L,L]$, with $y$ the coordinate in the extra dimension.
The UV and the IR boundaries are at orbifold fixed points $y=0$ and $y=L$, respectively.
We assume the inflaton $\phi$ and $\sigma$ to be localized on the UV and the IR boundaries, respectively.
A bulk Goldberger-Wise (GW) field~\cite{Goldberger:1999uk} $\Phi$ stabilizes the extra dimension.
The 5D action is then the sum of the gravitational and scalar field actions,
\es{}{
S &= \int \D^5x \, \sqrt{-G} (M_5^3 \mathcal{R}_5 - \Lambda_5 ) + S_\Phi + S_{\phi} + S_{\sigma},
}
with $M_5$ and ${\cal R}_5$ the 5D Planck scale and Ricci scalar, respectively, $\Lambda_5 = -12 M_5^3 k^2$ the 5D cosmological constant, and $k$ the $\text{AdS}_5$ curvature scale.
The 5D metric
\es{}{
\D s_5^2 = G_{AB}\D x^A \D x^B = n(y)^2 g_{\alpha\beta}\D x^\alpha \D x^\beta + \D y^2,
}
is determined by solving the Einstein equations~\cite{DeWolfe:1999cp}.
In particular, in the absence of the GW field, for $V_0 = 12 M_5^3 k$ and $V_L = -12 M_5^3 k$, one obtains the static RS1 geometry with $n(y) = \exp(-ky)$ and $g_{\alpha\beta} = \eta_{\alpha\beta}$, the Minkowski metric~\cite{Randall:1999ee}.
On the other hand, for $V_0 = 12 M_5^3 \sqrt{k^2 + H^2}$, we get the desired inflating 4D slices: $\D s_5^2 = n(y)^2 \D s_4^2 + \D y^2$, with $n(y) = \cosh(k y) - \sqrt{k^2 + H^2}\sinh(ky)/k$~\cite{Kaloper:1999sm, Nihei:1999mt, Kim:1999ja}.
In practice, in the regime of our interest where $m_{\rm KK}\simeq H$, the backreaction of the GW field on the 5D geometry is important, and the coupled Einstein and 5D Klein-Gordon equations need to be solved numerically.
We carry this out in Appendix~\ref{sec:app5D}.

With $n(y)$ at hand, we write the radion $r(x)$ and the spin-2 fluctuations $h_{\mu\nu}(x,y)$, satisfying $\nabla_\mu h^{\mu\nu}=h^\mu_\mu=0$,
\es{eq:5dmetric}{
\D s_5^2 = n^2\left(\frac{\pi ry}{L}\right)\D s_4^2 + \left(\frac{\pi r}{L}\right)^2\D y^2 + h_{\mu\nu}\D x^\mu \D x^\nu.
}
The VEV of the radion field $\langle r\rangle$ determines the `radius' of the extra dimension $\langle r\rangle = L/\pi $.
To derive the KK graviton couplings, we perform a KK decomposition $h_{\mu\nu}(x,y) = \sum_\ell n^2(y) \chi_\ell(y)\tilde{h}_{\mu\nu,\ell}(x)$, where $\tilde{h}_{\mu\nu,\ell}(x)$ are the various KK modes for different values of $\ell$ and $\chi_\ell(y)$ are the associated KK graviton wave functions in the extra dimension.
Each KK graviton mode $\tilde{h}_{\mu\nu,\ell}$, for $\ell\geq 1$, satisfies the equation of motion appropriate to a massive spin-2 particle in dS.
With the above parametrization, the action for the radion can also be derived.
The canonically normalized radion field is given by $\varphi(x) = F_\varphi \exp(-k\pi r(x))$~\cite{Kumar:2025anx}, where $F_\varphi$ depends on $n(y)$ and is computed in Appendix~\ref{sec:app5D}.

The coupling of $\sigma$ with $\tilde{h}_{\mu\nu,\ell}$ and radion fluctuations, $\delta\varphi = F_\varphi \exp(-k L)(-k\pi \delta r)$, follow from the kinetic term of $\sigma$, which is localized on the IR boundary,
\es{}{
&-\frac{1}{2}\sqrt{-G}\, G_{AB}\nabla^{A}\sigma \nabla^{B}\sigma \\
&= -\frac{1}{2}n^{2}(\pi r)\sqrt{-\bar{g}} \left(\bar{g}_{\mu\nu}+ \sum_{\ell}\chi_{\ell}\,\tilde{h}_{\mu\nu,\ell} \right) \, \nabla^{\mu}\sigma \nabla^{\nu}\sigma.
}
We expand $n^2(\pi r) \approx n^2(L) + 2 n(L)(\D n(L)/\D y) \pi \delta r$, and canonically normalize $\sigma\rightarrow \sigma/n(L)$, to obtain the radion-$\sigma$ coupling.
For the KK graviton coupling, we also canonically normalize $\tilde{h}_{\mu\nu,\ell} \rightarrow \tilde{h}_{\mu\nu,\ell}/\mpl$ so that the zero mode $\tilde{h}_{\mu\nu,0}$ couples to $\sigma$ suppressed by the 4D Planck scale $\mpl$.
These steps lead to the desired couplings,
\es{eq:int5D}{
\sqrt{-\overline{g}}\left({n'(L) \over k n(L)}{\delta\varphi \over \langle\varphi\rangle}\bar{g}_{\mu\nu} - \sum_{\ell=1} {\chi_\ell \tilde{h}_{\mu\nu,\ell}\over 2\mpl} \right)\nabla^{\mu}\sigma \nabla^{\nu}\sigma.
}
Comparing with Eq.~\eqref{eq:int}, we get $c_g/\Lambda_c = \chi_1(L)/\mpl$, for the lightest KK graviton, and $c_r/\Lambda_c = -2n'(L)/(k n(L)\langle\varphi\rangle)$.
By explicitly solving for the background 5D geometry and KK graviton mode functions, we evaluate these quantities in Appendix~\ref{sec:app5D}, finding $c_g/\Lambda_c \approx -0.24/H$ and $c_r/\Lambda_c \approx 0.04/H$.

\section{Interpretation of the Result and Observational Targets}
Having derived the radion and KK graviton couplings, we can compute $\fnl$ via Eq.~\eqref{eq:fnl}, once
we quantify the relation $\zeta(\vec{k}) = f \delta\sigma(\vec{k})$ discussed earlier.
To this end, we focus on the modulated reheating scenario~\cite{Dvali:2003em}.
In this scenario, the perturbations of inflaton $\phi$, and hence $\zeta$ generated during inflation, are negligible.
However, the decay rate $\Gamma$ of $\phi$ is dependent on another light field $\sigma$, which obtains large-scale fluctuations during inflation.
Thus, $\Gamma$ also fluctuates on large scales and different parts of the Universe get reheated at different times.
These modulations of the surface of reheating generate $\zeta$.

To make this explicit, we consider couplings $y\phi q q^c$ and $y\phi \sigma q q^c/M$, where $q, q^c$ are fermionic decay products of $\phi$ and $M$ is some UV scale.
In our 5D construction, these couplings can originate via some heavy bulk mediator with mass $\sim M$ that connect the IR-localized $\sigma$ with the UV-localized $\phi$.
These couplings give $\Gamma \propto y^2 m_\phi (1+\sigma/M)^2$, with $m_\phi$ the inflaton mass.
Owing to fluctuations $\delta\sigma(t, \vec{x})$, $\Gamma$ becomes location dependent.
In the limit when $\sigma_0\ll M$ and to the linear order, $\delta\Gamma/\Gamma \approx 2 \delta\sigma/M$.
To relate $\zeta$ to $\delta \Gamma/\Gamma$, we model the end of inflation through $\phi$ decay.
We solve the associated Boltzmann equations in Appendix~\ref{app:boltzmann}, finding $\zeta \approx (-1/6)\delta \Gamma/\Gamma$, consistent with~\cite{Dvali:2003em}.
This gives the desired relation, $\zeta(\vec{k}) = -\delta\sigma(\vec{k})/(3M)$.
The power spectrum $A_s$ then follows from the two point function $\langle\delta\sigma(\vec{k})\delta\sigma(-\vec{k})\rangle' = H^2/(2k^3)$ and is given by $A_s = H^2/(36\pi^2 M^2)$.
The observed value of $A_s \approx 2.1\times 10^{-9}$~\cite{SPT-3G:2025bzu} fixes $M\approx 1158 H$.
Using this and the shape function, $S = k_1^2 k_2^2 k_3^2 B(k_1, k_2, k_3)/{\rm Max} [|k_1^2 k_2^2 k_3^2 B(k_1, k_2, k_3)|]$, we rewrite Eq.~\eqref{eq:fnl} as,
\es{eq:fnl_simple_1}{
\fnl  = -{10 M \over 3 H^4} S_{\rm max}.
}
Here $S_{\rm max} = {\rm Max} [|k_1^2 k_2^2 k_3^2 B(k_1, k_2, k_3)|]$ and it depends on the massive particle mass $\mu$, along with $c_g, c_r, \Lambda_c, \dot{\sigma}_0$ that appear in Eqs.~\eqref{eq: quadratic mixing int} and~\eqref{eq: cubic mixing}.
To isolate these parametric dependencies, we rewrite Eq.~\eqref{eq:fnl_simple_1} as,
\begin{equation*}
\fnl^{\rm rad} = -{10 M \over 3 H^4} {c_r^2 \dot{\sigma}_0 \over \Lambda_c^2}S_{\rm max},~
\fnl^{\rm KK} = -{10 M \over 3 H^4} {c_g^2 \dot{\sigma}_0 \over 2\Lambda_c^2}S_{\rm max},
\end{equation*}
where $S_{\rm max}$ is computed assuming $c_g,~c_r=1$, $\Lambda_c=H$, $\dot{\sigma}_0=H^2$, and depend only on $\mu$.
\begin{figure}
    \centering
    \includegraphics[width=\linewidth]{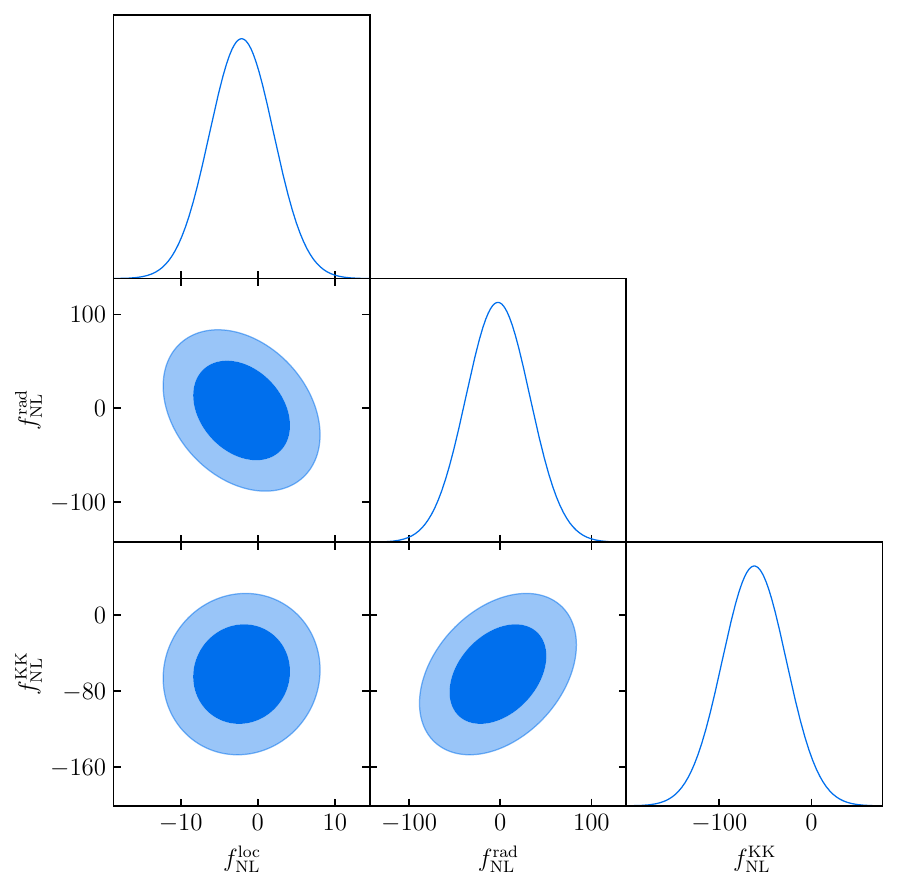}
    \caption{Triangle plot illustrating a simultaneous search for local NG, radion-mediated NG ($m_{\rm rad}\approx 1.80H$), and KK graviton-mediated NG ($m_{\rm KK} \approx 1.57H$). We include only the lightest KK graviton mode since that gives the largest NG.}
    \label{fig:triangle}
\end{figure}

As a benchmark, we consider the 5D parameters described in Appendix~\ref{sec:app5D} that gives $m_{\rm rad} \approx 1.80 H$, $m_{\rm KK}\approx 1.57 H$ and $\Lambda_c \approx 1.5H$, for which $S_{\rm max} \approx 0.27$ and $1.2$, respectively.
Plugging in values of $c_g/\Lambda_c$ and $c_r/\Lambda_c$ found earlier, and choosing $\dot{\sigma}_0/H^2=0.5$, we find $\fnl^{\rm rad} \approx -1$ and $\fnl^{\rm KK} \approx -66$.
These example values are consistent with our derived {\it Planck} bounds for these masses: $\fnl^{\rm rad} \approx 15\pm 34$ and $\fnl^{\rm KK} \approx -58\pm 32$, while expected to be better probed with future LSS data.
This choice of $\dot{\sigma}_0$ ensures that a derivative expansion in $|\dot{\sigma}_0|/ \Lambda_c^2$ $(\approx 0.1)$ is under control.
Furthermore, upon setting both $\sigma$ to their VEVs in Eq.~\eqref{eq:int}, we get tadpole terms for the KK graviton and the radion.
These tadpoles can be removed by doing a field shift as shown in Appendix~\ref{app: EFTofInf argument for spin-s}, which however modifies the $\sigma$ kinetic term. 
Our benchmark choices here ensure that these corrections are negligible.

The modulated reheating scenario itself predicts a local NG, $\fnl^{\rm loc}\gtrsim 1$~\cite{Zaldarriaga:2003my}, in addition to the cosmological collider signals.
Furthermore, the KK graviton exchanges also lead to contact terms~\cite{Cheung:2025dmc}, which have a more than $99\%$ overlap with the local shape, for the couplings we consider (Appendix~\ref{app:shape_compare}).
Therefore, for consistency, we search for all the three shapes: local, radion, (non-contact part of) KK graviton, simultaneously, and show the results in Fig.~\ref{fig:triangle}.
We find $\fnl=0$ at $95\%$ CL.
The correlations between local and radion, local and KK graviton, and radion and KK graviton shapes are $0.34$, $-0.18$, and $-0.39$, respectively, illustrating significant differences among the different shapes.

\section{Discussion}\label{sec: discussion}
In this work, we have performed the first observational search for extra-dimensional states: the radion modulus and KK gravitons, with the {\it Planck} 2018 temperature and polarization data.
Our search is consistent with $\fnl=0$ at 95\% CL, with the maximum local significance of $1.8\sigma$ for $m_{\rm KK}\approx 1.6 H$.
Our search differs qualitatively from previous searches for generic spin-2 particles, which focused on dimension-6 interactions, whereas KK graviton-mediated NG arises already through dimension-5 operators.
It would be interesting to search for the radion and KK gravitons both using the CMB trispectrum (four-point function) and especially future LSS data, which are expected to provide significantly improved sensitivity.
For the bispectra considered here, our sensitivity to oscillatory NG signals decreases when $m_{\rm KK}\gg H$ because vacuum fluctuations cannot efficiently produce these heavier KK gravitons on shell.
It is worth constructing explicit mechanisms that evade this suppression and generate observable signals from heavier on-shell KK gravitons.
While we have explored one example 5D construction, it would be useful to explore other set ups with different stabilization mechanisms and brane configurations.
Our results demonstrate that cosmological observations already probe higher-dimensional physics, with future surveys potentially capable of revealing direct evidence for the higher-dimensional origin of our Universe.

{\bf Acknowledgments.} We thank Mark Hertzberg, Qianshu Lu, Rashmish Mishra, Alize Sucsuzer, Raman Sundrum, Zhong-Zhi Xianyu, and Yisong Zhang for discussions, and Mark Hertzberg, Alize Sucsuzer, and Zhong-Zhi Xianyu for useful comments on an earlier version of this paper. This work was performed in part at the Aspen Center for Physics, which is supported by National Science Foundation grant PHY-2210452. The
authors acknowledge the \href{https://it.tufts.edu/high-performance-computing}{Tufts University High Performance Compute Cluster} which was utilized for the research reported here.

\newpage
\bibliography{biblio}

\clearpage

\onecolumngrid
\appendix 
\begin{center}
    \textbf{\large Appendix}\\[.2cm]
\end{center}
\setcounter{section}{0}

\numberwithin{equation}{section}
\renewcommand{\theequation}{\thesection\arabic{equation}}

\numberwithin{figure}{section}
\renewcommand{\thefigure}{\thesection\arabic{figure}}

\section{Details on Mode Functions and In-In Propagators}\label{app sec: Details on Mode Functions}

In this appendix, we present the mode functions, helicity conventions, and the in-in (Schwinger-Keldysh) propagators used to compute the results in the main text, as well as more details and comparisons of the KK graviton, the radion, and the contact part of the KK graviton shape functions with respect to their squeezed limits and angular dependence. 

\subsection{Mode Functions for Scalars and KK-gravitons}\label{app subsec: Mode Functions for Scalars and KK-gravitons}
The mode function for a massless scalar field in de Sitter is well known~\cite{Bunch:1978yq} (see~\cite{Baumann:2009ds} for a review)
\begin{equation}
    f_k(\eta) = \frac{H}{\sqrt{2k^3}}(1+ik\eta)e^{-ik\eta}\label{eq: massless scalar mode function}.
\end{equation}
These appear in the quantized scalar field as $\sigma(\vec{k}, \eta) = a_{-\vec{k}}^\dagger f^*_k(\eta) + a_{\vec{k}} f_k(\eta)$, along with the creation and annihilation operators.
The corresponding bulk-to-boundary propagator in the limit that $\eta_0\rightarrow 0$ on the boundary is given by,
\begin{equation}
    G_a(k,\eta) = \frac{H^2}{2k^3}(1-ik\eta a)e^{+ik\eta a}\label{eq: massless scalar bulk to boundary propagator}
\end{equation}
where $a =\pm$ denoting time-ordering ($+$) or anti-time-ordering ($-$) operations that is explained in more detail below.

The computation of the massive spin-2 mode functions can be found in \cite{Lee:2016vti} which is a culmination of the work found in \cite{0a992498-b0aa-38af-b240-75ee70eb334a,26fe42d6-bd6e-37c0-b45d-8bb27c0f126e,Deser:1983mm,Garidi:2003ys,Singh:1974qz,Singh:1974rc,Zinoviev:2001dt,Deser:2003gw,Wigner:1939cj,Deser:2001pe}.
Since we are using the same normalization conventions as in \cite{Lee:2016vti}, we simply quote their results. 
The massive spin-2 particle of interest, the KK graviton, denoted as $\tilde{h}_{\mu\nu}$ suppressing the mode index $\ell$, obeys the transverse-traceless (TT) conditions, and can be written as a Fourier mode expansion into helicity eigenstates as $\tilde{h}_{\mu\nu} = \sum_{\lambda = -2}^2\tilde{h}_{\mu\nu}^{(\lambda)}$. 
In the TT-gauge, we denote the traceless part of the spatial tensor with a hat, $\hat{h}_{ij}$ such that $\hat{h}_{ij}=\tilde{h}_{ij}-\tilde{h}_{\eta\eta}\delta_{ij}/3$ where $\delta_{ij}$ is the usual Euclidean 3-metric.
The polarization vectors and tensors obey the following relations
\begin{align}
    \hat{k}_i\epsilon_i^0& = 1,\:\:\:\hat{k}_i\epsilon^{\pm 1}_i = 0,\:\:\:\epsilon_i^{\pm 1} = (\epsilon_i^{\mp 1})^*,\:\:\:\epsilon_i^{\pm 1}(\epsilon_i^{\pm 1})^* = 2,\label{eq: spin 1 polarization vector identities}\\
    \hat{k}_i\epsilon_{ij}^0 & = \epsilon_j^0,\:\:\: \hat{k}_i\epsilon_{ij}^{\pm 1} = \frac{3}{2}\epsilon_j^{\pm 1},\:\:\:k_i\epsilon_{ij}^{\pm 2} = 0,\:\:\:\epsilon_{ij}^{\pm 2} = (\epsilon_{ij}^{\mp 2})^*,\:\:\: \epsilon_{ij}^{\pm 2}(\epsilon_{ij}^{\pm 2})^* = 4,\label{eq: spin 2 polarization tensor identitites}
\end{align}
where $\hat{k}_i = \vec{k}_i/|\vec{k}_i|$.
The normalization choice for the spin-1 polarization vectors fixes the longitudinal polarization vector to be $\epsilon_i^0 (\hat{\vec{k}}) = \hat{k}_i$.
The transverse polarization vectors span the 2D space orthogonal to $\hat{k}_i$.
The spin-2 polarization tensor identities~\eqref{eq: spin 2 polarization tensor identitites} lead to the following 
\begin{equation}
    \epsilon_{ij}^0= \frac{3}{2}\left(\hat{k}_i\hat{k}_j - \frac{1}{3}\delta_{ij}\right),\:\:\:\epsilon_{ij}^{\pm 1} = \frac{3}{2}\left(\hat{k}_i\epsilon_j^{\pm 1} + \hat{k}_j\epsilon_i^{\pm 1}\right),\:\:\: \epsilon_{ij}^{\pm 2}(\hat{\mathbf{z}}) = 
    \begin{pmatrix}
        1&\pm i&0\\
        \pm i&-1&0\\
        0&0&0
    \end{pmatrix},
    \label{eq: helicity 0 and 1 spin-2 polarization identities}
\end{equation} 
where the $\epsilon_{ij}^{\pm2}$ tensor is for the $\hat{\mathbf{k}}=\hat{\mathbf{z}}$ direction.  
The KK graviton mode functions are the following. 
For helicity $\lambda = \pm2$ there is a single mode function coming from the spatial piece, $\hat{h}_{ij}$
\begin{equation}
    \hat{h}_{ij}^{\pm 2}(k,\eta) = e^{i\pi/4}e^{-\pi\mu/2}\frac{\sqrt{\pi k}}{2H}(-k\eta)^{-1/2}H_{i\mu}^{(1)}(-k\eta) \epsilon_{ij}^{\pm 2}\label{eq: spatial helicity pm2 mode func}.
\end{equation}
For helicity $\lambda = \pm 1$, we have the purely spatial and spatio-temporal contributions 
\begin{align}
    \hat{h}_{ij}^{\pm 1}(k,\eta) & = \frac{i}{2}e^{i\pi/4}e^{-\pi\mu/2}\sqrt{\frac{\pi k}{6}}\frac{(-k\eta)^{-1/2}}{H\sqrt{(1/4+\mu^2)(9/4+\mu^2)}}\left[k\eta(H_{i\mu+1}^{(1)}(-k\eta) - H_{i\mu-1}^{(1)}(-k\eta)) - 3H_{i\mu}^{(1)}(-k\eta)\right]\epsilon_{ij}^{\pm 1},\label{eq: spatial helicity pm1 mode func}\\
    \tilde{h}_{i\eta}^{\pm 1}(k,\eta) & = e^{i\pi/4}e^{-\pi\mu/2}\sqrt{\frac{\pi k}{18}}\frac{(-k\eta)^{1/2}}{H\sqrt{(9/4+\mu^2)}}H_{i\mu}^{(1)}(-k\eta)\epsilon_i^{\pm 1}\label{eq: off diagonal helicity pm1 mode func}.
\end{align}
Finally, for helicity $\lambda = 0$, we get contributions from purely spatial, spatio-temporal, and purely temporal pieces, 
\begin{align}
    \hat{h}_{ij}^{0}(k,\eta) = &  \frac{1}{12}e^{i\pi/4}e^{-\pi\mu/2}\sqrt{\frac{\pi k}{6}}\frac{(-k\eta)^{-1/2}}{H\sqrt{(1/4+\mu^2)(9/4+\mu^2)}}\biggm[ 6k\eta((2+i\mu)H_{i\mu+1}^{(1)}(-k\eta) - (2-i\mu)H_{i\mu-1}^{(1)}(-k\eta))\nonumber\\
    & \:\:\:\:\:\:\:\:\:\:\:\:\:\:\:\:\:\:\:\:\:\:\:\:\:\:\:\:\:\:\:\:\:- (9-8k^2\eta^2)H_{i\mu}^{(1)}(-k\eta) \biggm]\epsilon_{ij}^0,\label{eq: spatial helicity 0 mode func}\\
    \tilde{h}_{i\eta}^{0}(k,\eta) =& \frac{i}{2}e^{i\pi/4}e^{-\pi\mu/2}\sqrt{\frac{\pi k}{6}}\frac{(-k\eta)^{1/2}}{H\sqrt{(1/4+\mu^2)(9/4+\mu^2)}}\left[ k\eta(H_{i\mu+1}^{(1)}(-k\eta) - H_{i\mu-1}^{(1)}(-k\eta)) - H_{i\mu}^{(1)}(-k\eta) \right]\epsilon_i^0,\label{eq: off diagonal helicity 0 mode func}\\
    \tilde{h}_{\eta\eta}^{0}(k,\eta) = & e^{i\pi/4}e^{-\pi\mu/2}\sqrt{\frac{\pi k}{6}}\frac{1}{H\sqrt{(1/4+\mu^2)(9/4+\mu^2)}}(-k\eta)^{3/2}H_{i\mu}^{(1)}(-k\eta)\label{eq: spin-2 mode functions}.
\end{align}
In performing calculations, we use the complex conjugation relation for the Hankel function, which normally is seen as $[H_{i\mu}^{(1)}(z)]^* = H_{-i\mu}^{(2)}(z) = e^{\pi\mu}H_{i\mu}^{(2)}(z)$ where in the last step we used the reflection formula. But here, we encounter a shifted Hankel function $i\mu\rightarrow i\mu+n$ for which if $n\in\mathbb{Z}$, then the conjugation is simply $[H_{i\mu+n}^{(1)}(z)]^* = (-1)^n e^{\pi\mu}H_{i\mu-n}^{(2)}(z)$.
A note should be made about our viable mass range when discussing massive spin-2 particles. 
The benchmark masses we are using are in the \textit{principal series} of the unitary representations for de Sitter space, which is defined as \cite{Deser:2001pe,Higuchi:1986py,0a992498-b0aa-38af-b240-75ee70eb334a,26fe42d6-bd6e-37c0-b45d-8bb27c0f126e,Baumann:2018muz}
\begin{equation}
    \frac{m^2}{H^2}\geq \left(s - \frac{1}{2}\right)^2\label{eq: principal series definition}
\end{equation}
where $s$ denotes spin. For us, at $s = 2$ this corresponds to $m_\text{KK}/H\geq 3/2$, which also satisfies the Higuchi bound~\cite{Higuchi:1986py} for spin-2, $m^2 \geq 2H^2$.

\subsection{In-In Propagators and Correlators}\label{subsec: in-in propagators}

In order to compute the NG signals shown above, we utilize the in-in (Schwinger-Keldysh) formalism which computes the correlation functions for time evolving operators defined as (this formalism was developed in \cite{Schwinger:1960qe,Bakshi:1962dv,Jordan:1986ug,Calzetta:1986ey} and \cite{Maldacena:2002vr,Weinberg:2005vy})
\begin{equation}
    \langle\mathcal{O}(t)\rangle = \bra{0}\overline{T}e^{i\int_{-\infty^+}^{t_0}\D t'\:\hat{H}_\text{int}(t')}\hat{\mathcal{O}}(t)Te^{-i\int_{-\infty^-}^{t_0}\D t'\:\hat{H}_\text{int}(t')}\ket{0},\label{eq: in-in master formula}
\end{equation}
where $(\overline{T})T$ is the (anti)-time ordering operator and $\hat{H}_\text{int}$ is the interacting Hamiltonian. 
The Feynman rules for computing these involve considering all possible permutations of time-ordered diagrams for a given process with two classes of propagators: bulk-to-boundary and bulk-to-bulk. 
The bulk-to-boundary propagator for $\delta\sigma$ was already given in Eq.~\eqref{eq: massless scalar bulk to boundary propagator}, while the four bulk-to-bulk propagators for the radion process are
\begin{align}
    D_{-+}(k,\eta_1,\eta_2) & = f_k(\eta_1)f_k(\eta_2)^*,\\
    D_{+-}(k,\eta_1,\eta_2) & = f_k(\eta_1)^* f_k(\eta_2),\\
    D_{++}(k,\eta_1,\eta_2) & = D_{-+}(k,\eta_1,\eta_2)\theta(\eta_1-\eta_2) + D_{+-}(k,\eta_1,\eta_2)\theta(\eta_2-\eta_1),\\
    D_{--}(k,\eta_1,\eta_2) & = D_{+-}(k,\eta_1,\eta_2)\theta(\eta_1-\eta_2) + D_{-+}(k,\eta_1,\eta_2)\theta(\eta_2-\eta_1).\label{eq: radion bulk to bulk propagators}
\end{align}
The bulk-to-bulk propagators for the KK graviton exchanges can be written in a similar manner by replacing the $f_k(\eta)$ mode functions with $\tilde{h}^{(\lambda)}_{\mu\nu}(k,\eta)$, but since we now have non-dynamical modes, such as $\tilde{h}_{\eta\eta}^0$, we should write them in terms of the \textit{dynamical} modes.
This generates additional contact terms in the in-in propagators \cite{Cheung:2025dmc}.\footnote{We thank Zhong-Zhi Xianyu for discussions on this point.} 
These contact terms lead to ``effective'' propagators
\begin{align}
    \widetilde{D}_{\pm\pm}^{(\eta\eta|\eta\eta)}(k,\eta_1,\eta_2) & = \widetilde{D}_{-+}^{(\eta\eta|\eta\eta)}(k,\eta_1,\eta_2)\theta(\eta_1-\eta_2) + \widetilde{D}_{+-}(k,\eta_1,\eta_2)^{(\eta\eta|\eta\eta)}\theta(\eta_2-\eta_1) \pm \Delta^{(\eta\eta|\eta\eta)}(k,\eta_1,\eta_2),\\
    \widetilde{D}_{\pm\pm}^{(i\eta|\eta\eta)}(k,\eta_1,\eta_2) & = \widetilde{D}_{-+}^{(i\eta|\eta\eta)}(k,\eta_1,\eta_2)\theta(\eta_1-\eta_2) + \widetilde{D}_{+-}(k,\eta_1,\eta_2)^{(i\eta|\eta\eta)}\theta(\eta_2-\eta_1) \pm \Delta^{(i\eta|\eta\eta)}(k,\eta_1,\eta_2),\\
    \widetilde{D}_{\pm\pm}^{(ij|\eta\eta)}(k,\eta_1,\eta_2) & = \widetilde{D}_{-+}^{(ij|\eta\eta)}(k,\eta_1,\eta_2)\theta(\eta_1-\eta_2) + \widetilde{D}_{+-}(k,\eta_1,\eta_2)^{(ij|\eta\eta)}\theta(\eta_2-\eta_1) \pm \Delta^{(ij|\eta\eta)}(k,\eta_1,\eta_2),\label{eq: effective KK propagators}
\end{align}
where the contact corrections are only present for $\pm\pm$ propagators, and not for the $+-$ or $-+$ propagators.
We have also used the notation $\widetilde{D}_{-+}^{(ij|\eta\eta)}(k,\eta_1,\eta_2) = \hat{h}_{ij}^{(0)}(k,\eta_1)[\tilde{h}_{\eta\eta}^{(0)}(k,\eta_2)]^*$, and so on.

As an example of how these contact terms arise, consider the
$(ij|\eta\eta)$ propagator. There are two possible sources of local
terms. The first comes from acting with the on-shell constraint
operator on the time-ordered propagator. This produces a delta-function
term from differentiating the step functions, but, as we shall see shortly, for
$(ij|\eta\eta)$ this term vanishes on the support of the delta function.
The second source is genuinely off shell since it comes from solving the
non-dynamical helicity-zero components at the level of the action in the presence of sources. This will be the contribution that
survives, and will give the contact term in Eq.~\eqref{eq: ij eta eta contact piece}.
For the case of $(ij|\eta\eta)$, we first start with (where the mode functions have now been stripped of their polarization tensors)
\begin{align}
    \widetilde{D}_{++}^{(ij|ij)}(k,\eta_1,\eta_2) = \hat{h}_{ij}^0(\eta_1)\hat{h}_{ij}^0(\eta_2)^*\theta(\eta_1-\eta_2) + \hat{h}_{ij}^0(\eta_1)^*\hat{h}_{ij}^0(\eta_2)\theta(\eta_2-\eta_1).
\end{align}
Now, recall that the transverse conditions are \cite{Kumar:2018jxz,Lee:2016vti}
\begin{equation}
    \tilde{h}_{i\eta}^0 = -{i\over k}\left(\tilde{h}_{\eta\eta}^{0}{}' - \frac{2}{\eta}\tilde{h}_{\eta\eta}^0\right)\:\:\:\text{and}\:\:\: \hat{h}_{ij}^0 = -{i\over k}\left(\tilde{h}_{i\eta}^0{}' - \frac{2}{\eta}\tilde{h}_{i\eta}^0\right) - \frac{1}{3}\tilde{h}_{\eta\eta}^0\label{eq: transverse constraints}.
\end{equation}
There is also the equation of motion for $\tilde{h}_{\eta\eta}^0$ that we can use, $\eta^2 \tilde{h}_{\eta\eta}^0{''} - 2\eta  \tilde{h}_{\eta\eta}^0{'} + (\eta^2k^2+m^2/H^2) \tilde{h}_{\eta\eta}^0 = 0$, to solve for $\tilde{h}_{\eta\eta}^0$ in terms of $\hat{h}_{ij}^0$, namely
\begin{align}
    \tilde{h}_{\eta\eta}^0 & = \frac{3k^2\eta^2}{N(\eta)}\left(2k^2\eta^2+\frac{3m^2}{H^2}-12-6\eta\partial_\eta\right)\hat{h}_{ij}^0\nonumber\\
    &\equiv \mathcal{D}(\eta)\hat{h}_{ij}^0\label{eq: operator for going from ij to eta eta component}
\end{align}
where $N(\eta)\equiv 4k^4\eta^4 + 12(m^2/H^2-2)k^2\eta^2+9m^2(m^2/H^2-2)/H^2$. We can use the operator $\mathcal{D}(\eta)$ to generate $\widetilde{D}_{++}^{(ij|\eta\eta)}$ in the following manner
\begin{align}
    \widetilde{D}_{++}^{(ij|\eta\eta)}(k,\eta_1,\eta_2) & = \mathcal{D}(\eta_2)\widetilde{D}_{++}^{(ij|ij)}(k,\eta_1,\eta_2)\nonumber\\
    & = \widetilde{D}_{-+}^{(ij|\eta\eta)}\theta_{12} + \widetilde{D}_{+-}^{(ij|\eta\eta)}\theta_{21} + \frac{18k^2\eta_2^3}{N(\eta_2)}\left(\hat{h}_{ij}^0(\eta_1)^*\hat{h}_{ij}^0(\eta_2) - \hat{h}_{ij}^0(\eta_1)\hat{h}_{ij}^0(\eta_2)^*\right)\delta(\eta_1-\eta_2)
\end{align}
where $\theta_{12}\equiv \theta(\eta_1-\eta_2)$ for shorthand. This last term has a delta function with support at $\eta_1=\eta_2$, and thus vanishes. We would then naively conclude that there is no contact term. In fact, to generate a contact term at this level, we would need an operator of the form in \eqref{eq: operator for going from ij to eta eta component} but with second order derivatives in $\eta_1$ and/or $\eta_2$ to produce a Wronskian. This should not be surprising as these are the on-shell constraints. But there is another contribution for contact terms at the level of the action, namely from integrating out the non-dynamical modes with the inclusion of source terms for the dynamical and non-dynamical modes. In doing so, we can derive and read off from the action the equations of motion as an operator relation between $\tilde{h}_{\eta\eta}^0$ and $\hat{h}_{ij}^0$ of the form $\tilde{h}_{\eta\eta}^0 = \mathcal{O}(\eta)\hat{h}_{ij}^0$ where  $\mathcal{O}(\eta) = \mathcal{D}(\eta) + \lambda(\eta)\mathcal{E}(\eta)$ with $\lambda(\eta) = 4H^2k^2\eta^2\alpha_1\alpha_2/9$ where $\alpha_1 \equiv (H^2(\mu^2+9/4))^{-1}$ and $\alpha_2\equiv (H^2(\mu^2+1/4))^{-1}$. 
The coefficient $\lambda(\eta)$ is not fixed by the on-shell operator
$\mathcal D(\eta)$. It is read off from the off-shell constraint
solution obtained after integrating out the non-dynamical components at
the level of the action. The operator $\mathcal E(\eta)$ is
the equation-of-motion in operator form for the dynamical traceless spatial mode
$\hat h^0_{ij}$. Thus, $\mathcal E\hat h^0_{ij}=0$ for the free mode
function, but the propagator obeys
\begin{equation}
    \mathcal{E}(\eta_2)\widetilde{D}_{++}^{(ij|ij)}(k,\eta_1,\eta_2) = i\delta(\eta_1-\eta_2).\label{eq: propagator eom} 
\end{equation}
As we saw above, when we operate with $\mathcal{O}(\eta_2)$ on $\widetilde{D}^{(ij|ij)}_{++}$ to produce $\widetilde{D}^{(ij|\eta\eta)}_{++}$, the first term $\mathcal{D}(\eta_2)$ produces no contact term, but this last term will thanks to~\eqref{eq: propagator eom}. This is what leads to the correction in~\eqref{eq: ij eta eta contact piece}. The rest of them are the following~\cite{Cheung:2025dmc}
\begin{align}
    \Delta^{(\eta\eta|\eta\eta)}(k,\eta_1,\eta_2) & = -\frac{2iH^2\alpha_1\alpha_2}{3}\partial_{\eta_1}\left(\eta_1^2\partial_{\eta_1}\delta(\eta_1-\eta_2)\right) - \frac{2i\alpha_1}{3}\left(1 - H^2k^2\eta_1^2\alpha_2\right)\delta(\eta_1-\eta_2)\\
    \Delta^{(i\eta|\eta\eta)}(k,\eta_1,\eta_2) & = \frac{2iH^2k\alpha_1\alpha_2}{3}\partial_{\eta_2}\left(\eta_2^2\delta(\eta_1-\eta_2)\right)\\
    \Delta^{(ij|\eta\eta)}(k,\eta_1,\eta_2) & = \frac{4iH^2k^2\eta_2^2\alpha_1\alpha_2}{9}\delta(\eta_1-\eta_2).\label{eq: ij eta eta contact piece}
\end{align}
Note that the ordering of the contact terms differs from that in \cite{Cheung:2025dmc} but are related using $\Delta^{(ab|cd)}(k,\eta_1,\eta_2) = \Delta^{(cd|ab)}(k,\eta_2,\eta_1)$. 

To compute the in-in expectation value of $\hat{\mathcal{O}}(t) = \delta\sigma_1\delta\sigma_2\delta\sigma_3$, where $\delta\sigma_i\equiv \delta\sigma(\mathbf{k}_i)$, we have three different cubic vertices involving the KK graviton and two vertices involving the radion, as seen in Eqs.~\eqref{eq: quadratic mixing int} and~\eqref{eq: cubic mixing}. Note that we switch to conformal time ($\eta$) derivatives here to simplify calculations which is related to physical time ($t$) derivatives as $\partial_t\leftrightarrow (-H\eta)\partial_\eta$.   To keep the computation general, we denote the vertices as $V_a[G_a\cdots]$ that can act on the bulk-to-boundary propagators. These include taking appropriate (conformal) time and spatial derivatives of bulk-to-boundary propagators. For example, when exchanging the radion in \eqref{eq: radion contribution to the bispecturm} the $V_a$ vertex operators will contain two derivatives of either spatial or (conformal) time, while $V_b \propto \dot{\sigma}_0\eta_2^{-3}\partial_\eta(\cdot)$ up to a coupling constant. While for the exchange of the KK-graviton in \eqref{eq: KK-graviton contribution to bispectrum}, $V_a$ always contains two derivatives, either spatial, (conformal) time, or a mixture, while $V_b$ will actually contain no derivatives at all and is of the form $V_b\propto \dot{\sigma}_0\eta_2^{-2}$. 
The bispectrum then takes the form
\begin{align}
\langle\delta\sigma_1\delta\sigma_2\delta\sigma_3\rangle &= -ab\int_{-\infty}^0d\eta_1 d\eta_2\:V_a[G_a(k_1,\eta_1)G_a(k_2,\eta_1)]D_{ab}(k_3,\eta_1,\eta_2)V_b[G_b(k_3,\eta_2)]\nonumber\\
    & =  \EnergyInjectionBispectrum[0.45] + \EnergyInjectionBispectrumMM[0.45] + \EnergyInjectionBispectrumPM[0.45] + \EnergyInjectionBispectrumMP[0.45],
\end{align}
where we replace $D_{ab}\rightarrow \widetilde{D}_{ab}$ for exchanging the KK graviton, and the second line are the Feynman diagrams for all possible ordered diagrams in the $s$-channel for exchanging either the KK graviton or the radion. 
By a permutation of the external legs, we can generate the $t$- and $u$-channels as well.

The radion contribution to the bispectrum can then be written as
\begin{align}
    \langle \delta\sigma_1\delta\sigma_2\delta\sigma_3\rangle' = -\frac{c_r^2\dot{\sigma}_0}{\Lambda_c^2H^5}&\sum_{a,b = \pm}(-ab)\int_{-\infty}^0 \frac{\D\eta_1}{\eta_1^2}\int_{-\infty}^0\frac{\D\eta_2}{\eta_2^3}\nonumber\\
    &\times\left(G_a'(k_1,\eta_1) G_a'(k_2,\eta_1) + (\mathbf{k}_1\cdot\mathbf{k}_2)G_a(k_1,\eta_1)G_a(k_2,\eta_1)\right)D_{ab}(k_3,\eta_1,\eta_2) G_b'(k_3,\eta_2), \label{eq: radion contribution to the bispecturm}
\end{align}
where $G_a'(k,\eta)\equiv \partial_\eta G_a(k,\eta)$ and $\mathbf{k}_1\cdot\mathbf{k}_2 = (k_3^2-k_1^2-k_2^2)/2$ by momentum conservation. 

In the bispectrum contribution from exchanging the KK graviton, only the $\lambda = 0$ helicity contributes since only that can mix with scalar $\sigma$ fluctuations and appear in the the linear mixing vertex by conservation of angular momentum. 
In this case however, there are multiple terms coming from each component of the KK graviton which becomes
\begin{align}
    \langle\delta\sigma_1
    \delta\sigma_2\delta\sigma_3\rangle' = -\frac{c_g^2\dot{\sigma}_0}{\Lambda_c^2H}&\sum_{a,b=\pm}(-ab)\int_{-\infty}^0 \D\eta_1\int_{-\infty}^0\frac{\D\eta_2}{\eta_2^2}G_b(k_3,\eta_2)\times\biggm[ -G_a'(k_1,\eta_1)G_a'(k_2,\eta_1)\widetilde{D}_{ab}^{(\eta\eta|\eta\eta)}(k_3,\eta_1,\eta_2)\nonumber\\
    &  +i\left(\epsilon_i^0(\hat{k}_3)k_1^iG_a(k_1,\eta_1)G_a'(k_2,\eta_1) + \epsilon^0_i(\hat{k}_3)k_2^iG_a(k_2,\eta_1)G_a'(k_1,\eta_1)\right)\widetilde{D}_{ab}^{(i\eta|\eta\eta)}(k_3,\eta_1,\eta_2)\nonumber\\
    & - \epsilon_{ij}^0(\hat{k}_3)k_1^ik_2^jG_a(k_1,\eta_1)G_a(k_2,\eta_1)\widetilde{D}_{ab}^{(ij|\eta\eta)}(k_3,\eta_1,\eta_2)\biggm]\label{eq: KK-graviton contribution to bispectrum}.
\end{align}
Note that here we have written the polarization vectors/tensors explicitly to display their contributions in Eq.~\eqref{eq: KK-graviton contribution to bispectrum}. 
Their contractions are $\epsilon_i^0(\hat{k}_3)k^i_1 = (k_2^2-k_1^2-k_3^2)/2k_3$, $\epsilon_i^0(\hat{k}_3)k^i_2 = (k_1^2-k_2^2-k_3^2)/2k_3$, and $\epsilon_{ij}^0(\hat{k}_3)k_1^ik_2^j = (k_3^4+2k_3^2(k_1^2+k_2^2)-3(k_1^2-k_2^2)^2)/8k_3^2$, using Eqs.~\eqref{eq: spin 1 polarization vector identities},~\eqref{eq: spin 2 polarization tensor identitites}, and~\eqref{eq: helicity 0 and 1 spin-2 polarization identities}. For clarity, we also write out the form of the contact terms hidden within \eqref{eq: KK-graviton contribution to bispectrum} since they can be analytically derived
\begin{align}
    \langle\delta\sigma_1\delta\sigma_2\delta\sigma_3\rangle'\rvert_{\rm contact} & \supset -\frac{c_g^2\dot{\sigma}_0}{\Lambda_c^2H}\sum_{ab = \pm}(-ab)\int_{-\infty}^0 d\eta_1 \biggm[ \frac{2i\alpha_1}{3}\frac{G_b(k_3,\eta_1)}{\eta_1^2}\big[ H^2\alpha_2\partial_{\eta_1}\left(\eta_1^2\partial_{\eta_1}(G_a'(k_1,\eta_1)G_a'(k_2,\eta_1))\right)\nonumber\\
    & + (1-H^2k_3^2\alpha_2)G_a'(k_1,\eta_1)G_a'(k_2,\eta_1)\big] +\frac{4iH^2 k_3^2\alpha_1\alpha_2}{9} \epsilon_{ij}^0(\hat{k}_3)k^i_1k^j_2 G_b(k_3,\eta_1) G_a(k_1,\eta_1)G_a(k_2,\eta_1)\nonumber\\
    &+\frac{2H^2k_3\alpha_1\alpha_2}{3}\eta_1^2\partial_{\eta_1}\left(\frac{G_b(k_3,\eta_1)}{\eta_1^2}\right)\left[\epsilon_i^0(\hat{k}_3)k_1^iG_a(k_1,\eta_1)G_a'(k_2,\eta_1) + \epsilon_i^0(\hat{k}_3)k^i_2G_a(k_2,\eta_1)G_a'(k_1,\eta_1)\right]\biggm]\nonumber\\
    & =\left(-\frac{c_g^2\dot{\sigma}_0}{\Lambda_c^2H}\right)\frac{2\left(42k_T^4e_2-14k_T^6+\left(24\mu^2-100\right)k_T^3e_3+\left(29-12\mu^2\right)\left(k_T^2e_2^2+k_Te_2e_3\right)-192e_3^2\right)}{9k_T^3e_3^3(4\mu^2+1)(4\mu^2+9)
},\label{eq: total contact contribution}
\end{align}
where \eqref{eq: total contact contribution} includes all three channels and $k_T \equiv k_1+k_2+k_3$, $e_2\equiv k_1k_2 + k_1k_3 + k_2k_3$, and $e_3\equiv k_1k_2k_3$.

\subsection{Comparison between Shape Functions}\label{app:shape_compare}

It is well known that the squeezed limit bispectrum for massless scalars exchanging a massive particle with spin-$s$ is sensitive to the particle's spin: $\langle\delta\sigma_1\delta\sigma_2\delta\sigma_3\rangle\propto P_s(\cos\theta)F_s(k_1,k_2,k_3)$, where $P_s(x)$ is the Legendre polynomial and $F_s(k_1,k_2,k_3)$ is some function dependent upon the kinematics~\cite{Arkani-Hamed:2015bza, Lee:2016vti, Bordin:2018pca}.
Here, $\theta$ is the angle between the long and short momenta $\theta\equiv \cos^{-1}(\hat{\mathbf{k}}_1\cdot\hat{\mathbf{k}}_3)$.
The angular dependence here is similar to the angular dependence that shows up at terrestrial particle colliders where the differential cross section plotted against $\cos\theta$ displays a characteristic shape dependent upon the spin of the exchanged particle \cite{ATLAS:2016rnf}.

We can then parameterize one of the momenta as $k_2(\theta) = k_1\sqrt{1+x_3^2 + 2x_3\cos\theta}$ where $x_3\equiv k_3/k_1$. We also set $k_1 = 1$, such that the shape function $S$ is a function of only $x_3$ and $\theta$. 
The KK graviton angular dependence is shown below. 
\begin{figure}[H]
\centering
\subfloat[KK graviton shape function scaled by a factor of $3/5$.\label{fig:sub1}]{
  \includegraphics[width=0.48\columnwidth]{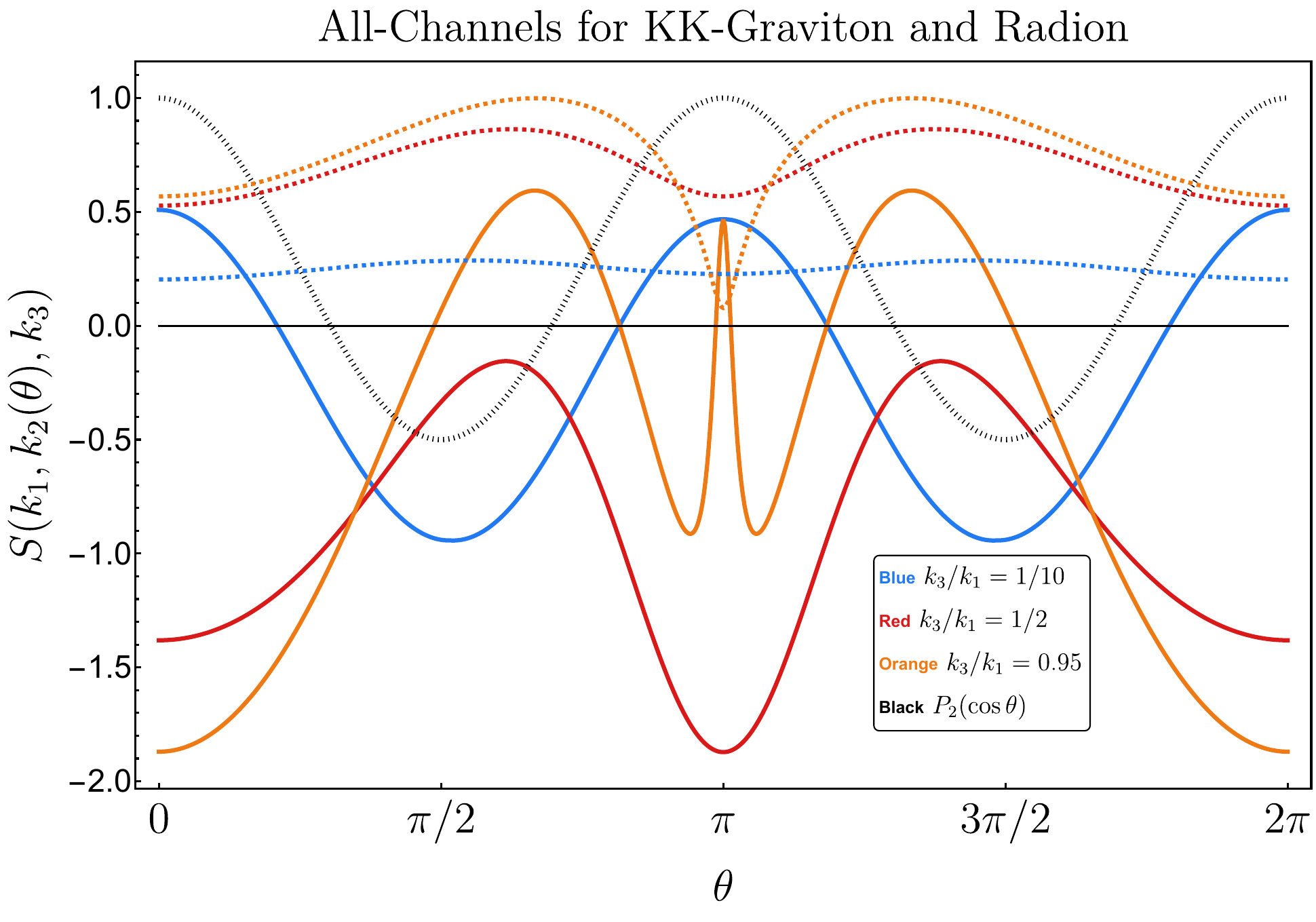}
}
\hfill
\subfloat[KK graviton shape scaled by $1/6$.\label{fig:sub2}]{
  \includegraphics[width=0.48\columnwidth]{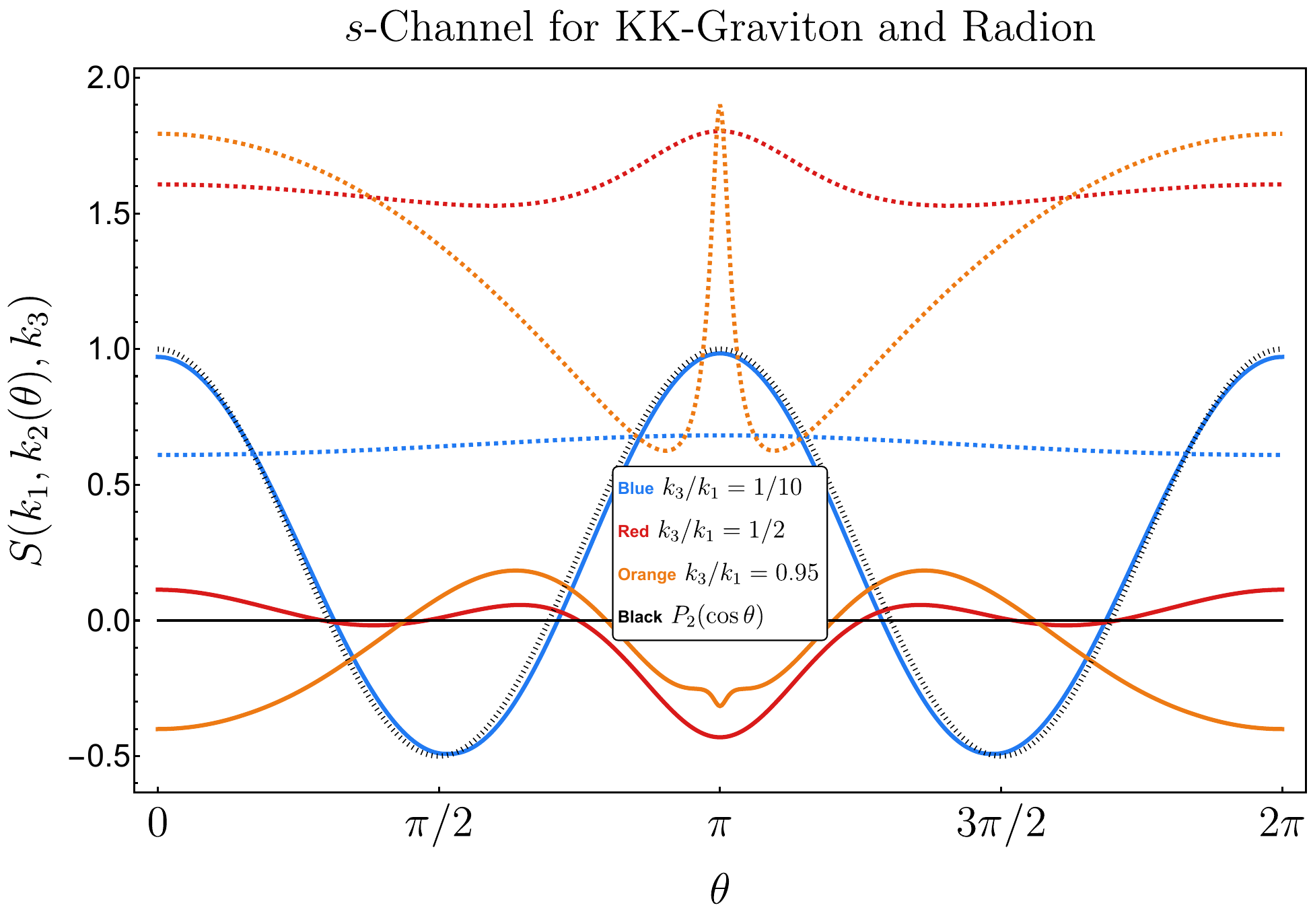}
}
\caption{The shape functions for the exchange of the KK graviton, excluding the contact contribution, (solid color curves) and the radion (dashed color curves), parameterized by $(k_3/k_1,\theta)$ with $\mu = 1.1$ for both KK graviton and radion as in Fig. \ref{fig:shape}. We compare these to $P_2(\cos\theta)$ shown in dashed black. As we take the squeezed limit, the KK graviton shape function approaches $P_2(\cos\theta)$, with an almost perfect match already at $k_3/k_1=0.1$. Notably, the radion shape functions do not switch signs, as a function of $\theta$.}
\label{fig: KK graviton angular dependence comparison}
\end{figure}

As can be seen in Fig.~\ref{fig: KK graviton angular dependence comparison}, the KK graviton (the solid curves) displays the expected angular behavior almost matching $P_2(\cos\theta)$ in the squeezed limit, $k_3/k_1\ll 1$. 
We have scaled the KK-graviton shape function by some numerical factor given in \ref{fig:sub1} and \ref{fig:sub2} in order to make the comparison with $P_2(\cos\theta)$ apparent.
This match is almost exact for the s-channel, while the addition of the $t$- and $u$-channels shifts the shape vertically.
However, we cannot scale the radion shape function to match any form of $P_s(\cos\theta)$.
This is also evident from the fact that the radion shape function never oscillates around zero.

As we start to deviate away from the squeezed limit, the angular structure changes drastically, especially when $k_3/k_1\sim 1$ at $\theta= \pi$. We can draw the momentum triangle to better understand what is occurring, along with the other configurations seen in \ref{fig: KK graviton angular dependence comparison}. Consider the case for when we are at $\theta = 2\pi/3$ for the following momentum triangles (note that the depicted angle is $\pi - \theta$)
\begin{align*}
    \MomentumTriangleVar[0.8]{0.1}{2*pi/3}&\MomentumTriangleVar[0.8]{0.5}{2*pi/3}
    \MomentumTriangleVar[0.8]{2/3}{2*pi/3}\MomentumTriangleVar[0.8]{0.95}{2*pi/3}
\end{align*}
We can see at a fixed angle, that as the ratio $k_3/k_1\rightarrow 1$, we approach the equilateral configuration.
However, for $k_3/k_1\rightarrow 1$, if we reduce the depicted angle, we approach $\theta\rightarrow \pi$, for which $k_2 \rightarrow 0$. The shape function, when all channels are included, then diverges at $\theta = \pi$, due to the dependence of the shape function on inverse powers of $k_2$. 

Now, when it comes to the radion, we would not expect any angular dependence in the form of $P_s(\cos\theta)$ since $P_0(\cos\theta) = 1$. However, there can be operators involving products of momentum such as $\mathbf{k}_1\cdot\mathbf{k}_3$ which, given our angle parameterization above, is $\mathbf{k}_1\cdot\mathbf{k}_3 = k_1k_3\cos\theta$. Therefore in the squeezed limit, we expect an almost straight line against $\theta$, and then deviations due to $k_2^{-1}$ behavior. The radion shape function is displayed in \ref{fig:sub1} and \ref{fig:sub2} as the dashed curves. 
We can clearly see some form of angular dependence, but this should not be mistaken for the exchange of a spinning particle since the shape profile in any limit does not reach any form of $P_s(\cos\theta)$. We can see however that in the squeezed limit of $k_3/k_1\ll 1$, we get the expected result of a straight line. This is even more true when looking at just the $s$-channel in \ref{fig:sub2} at slightly larger values of $k_3/k_1$. For illustrative purposes to support these graphs, consider the following momentum diagrams as $\theta$ changes for the ratio of $k_3/k_1 = 0.5$ which shows the approach of the soft limit (while maintaining proper order of the momentum's lengths):
\begin{align*}
    \MomentumTriangleVar[0.8]{0.5}{1.85}& \MomentumTriangleVar[0.8]{0.5}{2.2} \MomentumTriangleVar[0.8]{0.5}{2.75}\MomentumTriangleVar[0.8]{0.5}{3}
\end{align*}

Finally for completeness on this discussion of the angular dependence, we can plot how the KK graviton shape compares with the contact shape that is generated by the effective propagators with the bispectrum contribution given in \eqref{eq: total contact contribution}. Of course, we expect no contribution to the angular dependence in the squeezed limit, and we see such behavior.

\begin{figure}[H]
\centering
\subfloat[The KK graviton shape function scaled by a factor of $3/5$ while the contact piece has been scaled by $1/8$.\label{fig:sub3}]{
  \includegraphics[width=0.48\columnwidth]{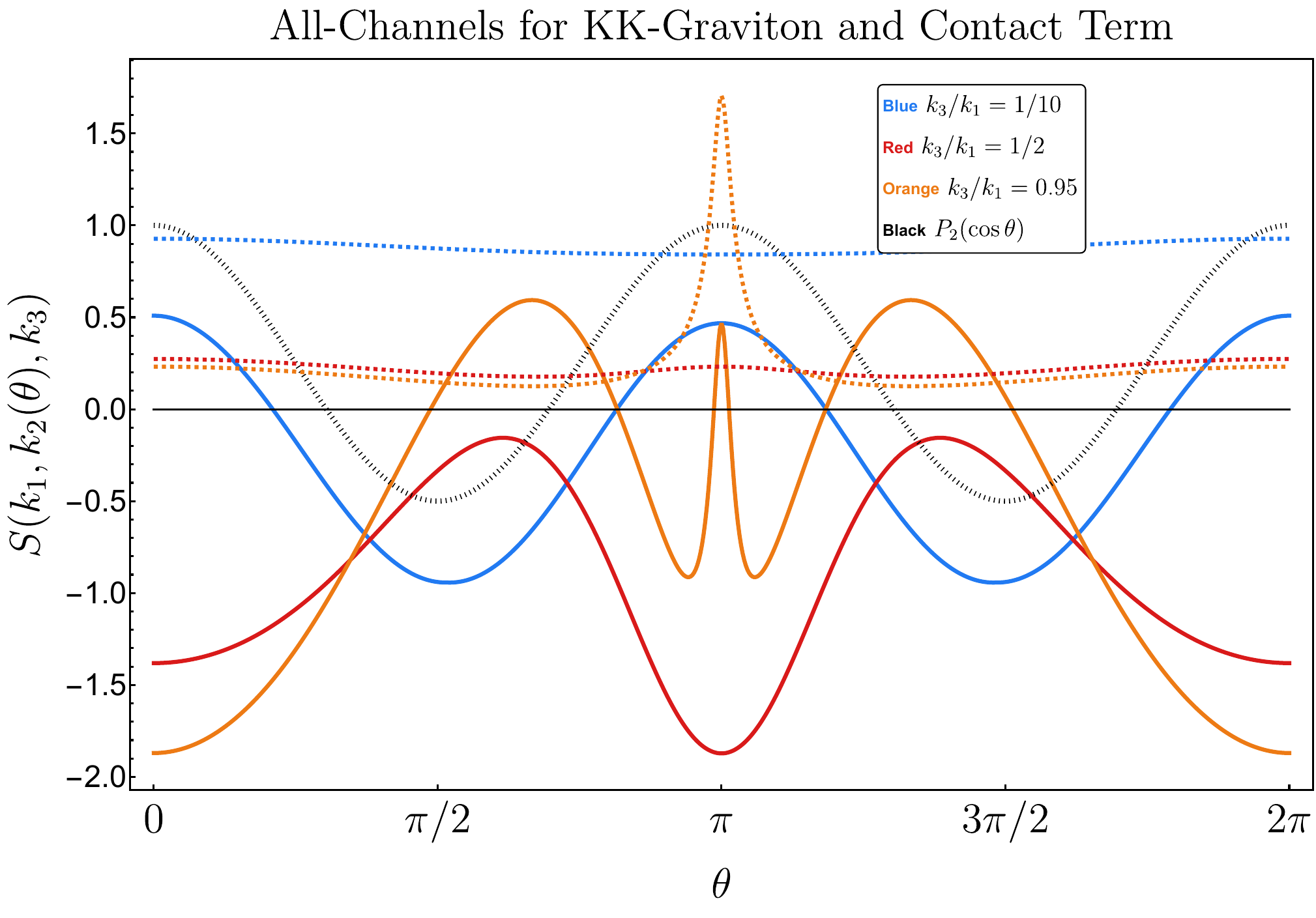}
}
\hfill
\subfloat[The KK graviton shape scaled by $1/6$ while the contact piece has been scaled by $1/16$.\label{fig:sub24}]{
  \includegraphics[width=0.48\columnwidth]{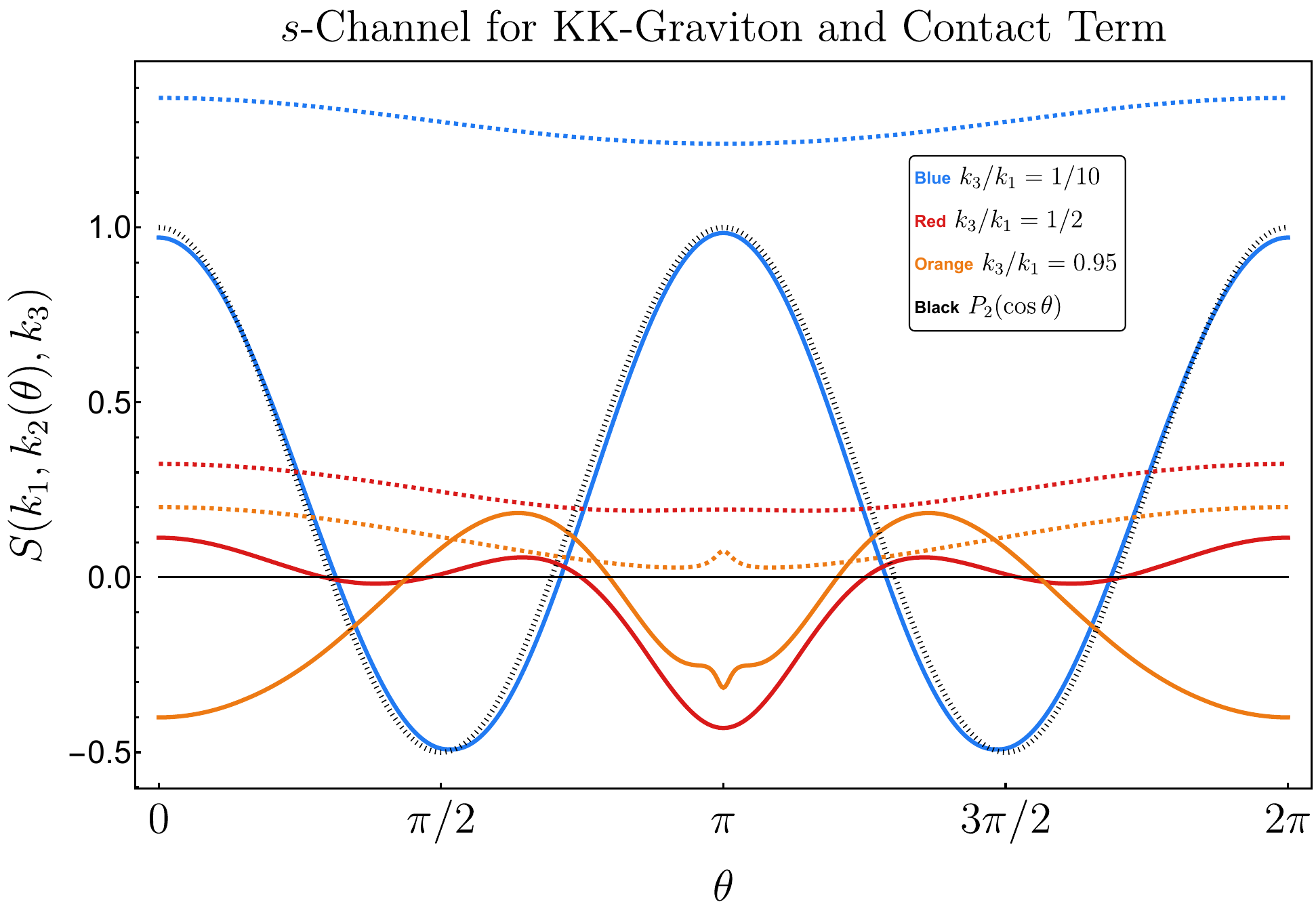}
}
\caption{The shape functions for the exchange of the KK graviton (solid color curves) and the contact term in \eqref{eq: total contact contribution} (dashed color curves) parameterized by $(k_3/k_1,\theta)$ with $\mu = 1.1$ for the KK graviton as in Fig. \ref{fig:shape}. We compare these to $P_2(\cos\theta)$ shown in dashed black. Notably, the contact shape functions do not switch signs as a function of $\theta$.}
\label{fig: KK graviton angular dependence comparison to the contact shape}
\end{figure}

It is interesting to determine how much overlap there is between this contact piece and the local shape function with corresponding bispectrum of the form $B(k_1,k_2,k_3) = (\sum_{i=1}^3k_i^3)/\prod_{i=1}^3k_i^3$. To this end, we can compare the two by their orthogonality as defined in \cite{Babich:2004gb}
\begin{equation}
    \cos(S_1,S_2) = \frac{\int_0^1dx_2\int_{1-x_2}^1 dx_3\: S_1S_2(1,x_2,x_3)}{\left[\int_0^1dx_2\int_{1-x_2}^1 dx_3\:S_1(1,x_2,x_3)^2\right]^{1/2}\left[\int_0^1 dx_2\int_{1-x_2}^1 dx_3\: S_2(1,x_2,x_3)^2\right]^{1/2}}
\end{equation}
where $S_1$ and $S_2$ are any two shape functions, $x_2 = k_2/k_1$, $x_3 = k_3/k_1$, and the physical domain is then $0\leq x_2\leq1$, $0\leq x_3\leq1$, and $x_2+x_3\geq1$. For the comparison between the contact and local shape function at mass value $\mu = 1.1$, we find  $\cos(S_\text{contact},S_\text{local}) = 0.99$, indicating a strong overlap between the two. We can perform the same calculation between $S_\text{KK-graviton}$ and $S_\text{contact}$ for $\mu = 0.5$ and $\mu = 1.1$ which we find to be the following $\cos(S_\text{KK-graviton},S_\text{contact}) = 0.13 $ for $\mu = 0.5$ and  $\cos(S_\text{KK-graviton},S_\text{contact}) = 0.42$ for $\mu = 1.1$.
We also compute shape correlations using {\tt CMB-BEST} that takes into account projection effects, and find a correlation of $0.99$ between $S_{\rm contact}$ and $S_{\rm local}$, and $0.20$ between $S_{\rm KK-graviton}$ and $S_{\rm contact}$, for for $\mu = 0.5$.
\begin{figure}[H]
\centering

\subfloat[All channels, $k_3/k_1=1/10$.\label{fig:radion-all-01}]{
  \includegraphics[width=0.47\textwidth]{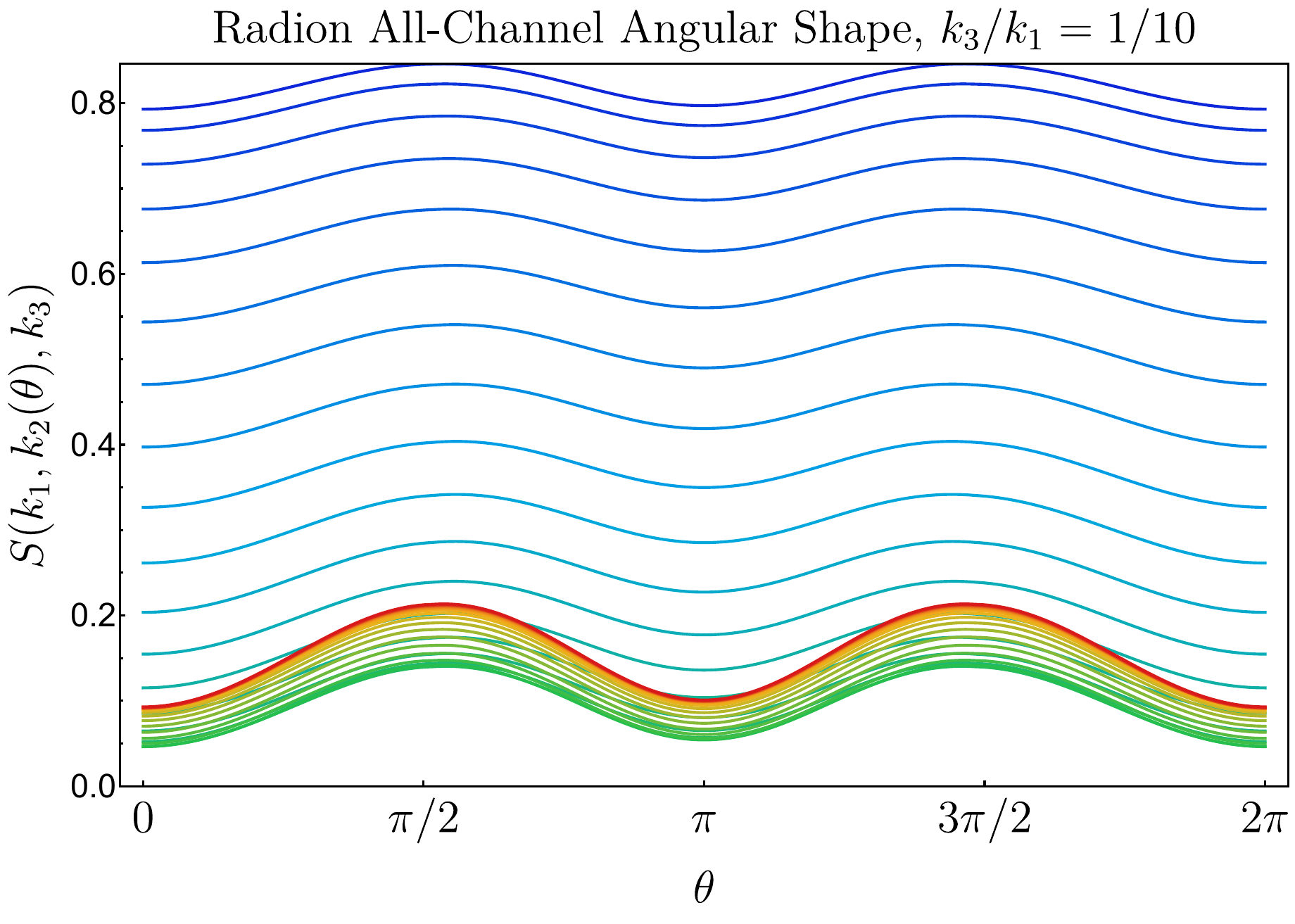}
}
\hfill
\subfloat[$s$-channel, $k_3/k_1=1/10$.\label{fig:radion-s-01}]{
  \includegraphics[width=0.47\textwidth]{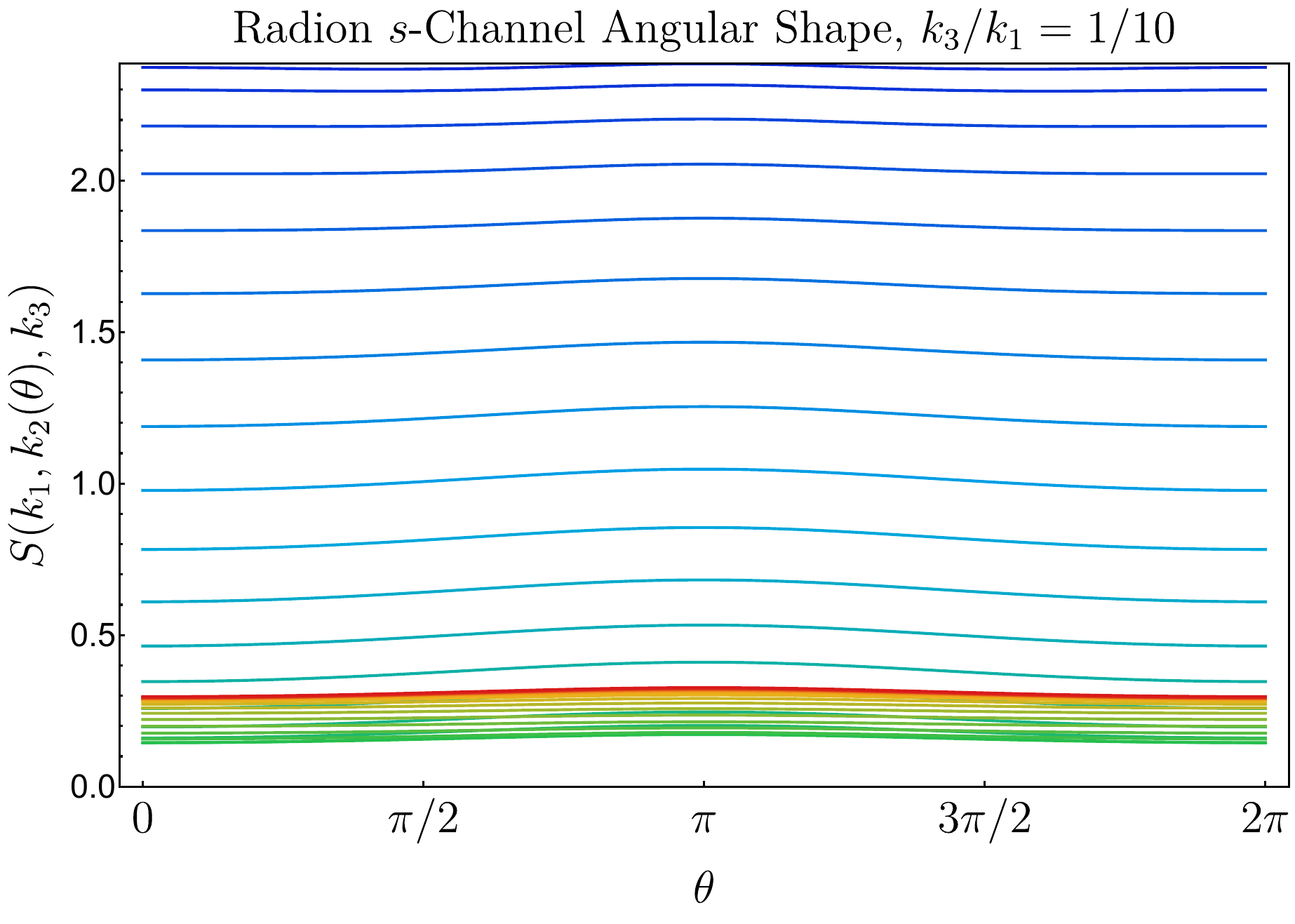}
}

\vspace{-0.6em}

\subfloat[All channels, $k_3/k_1=1/2$.\label{fig:radion-all-05}]{
  \includegraphics[width=0.47\textwidth]{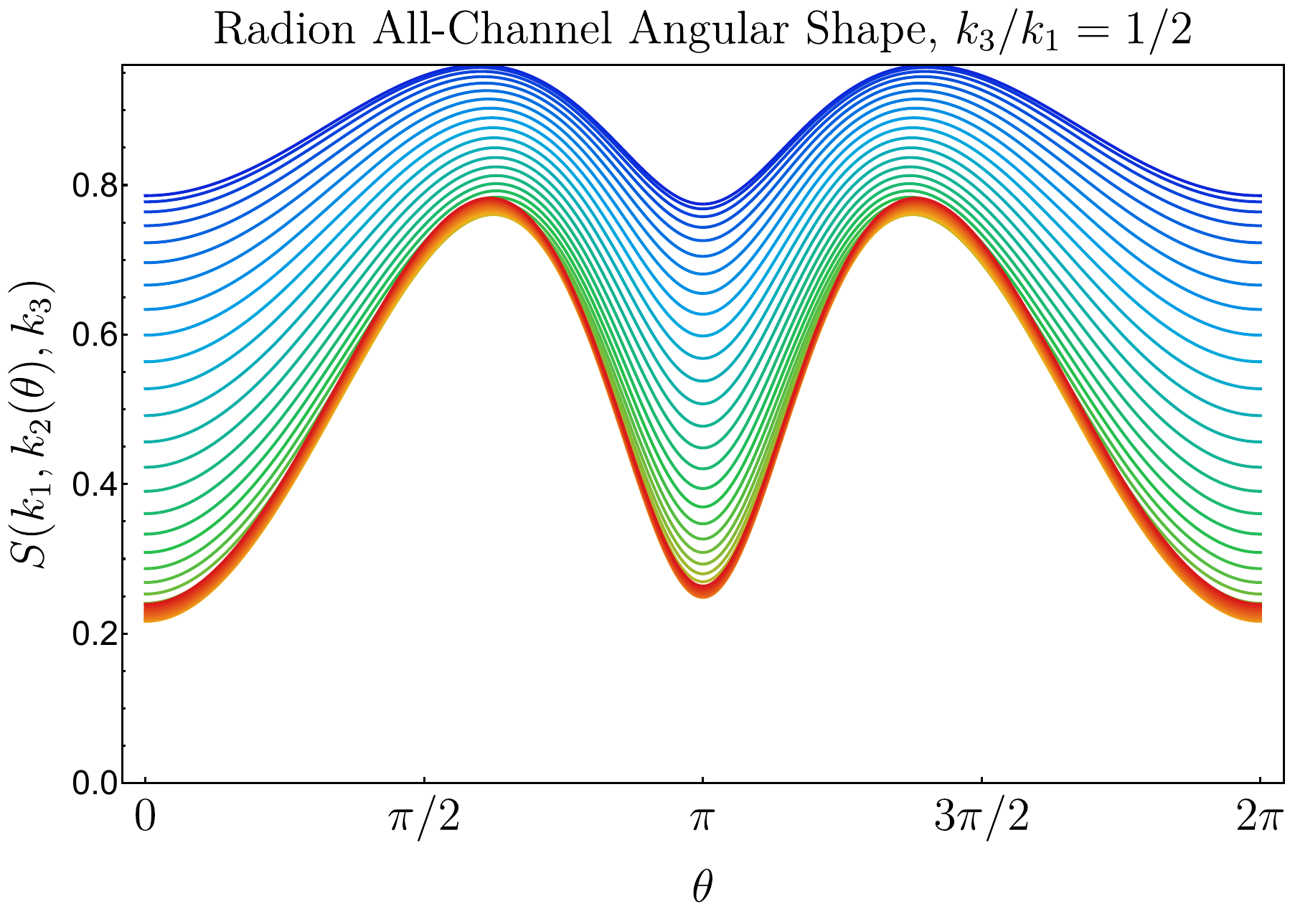}
}
\hfill
\subfloat[$s$-channel, $k_3/k_1=1/2$.\label{fig:radion-s-05}]{
  \includegraphics[width=0.47\textwidth]{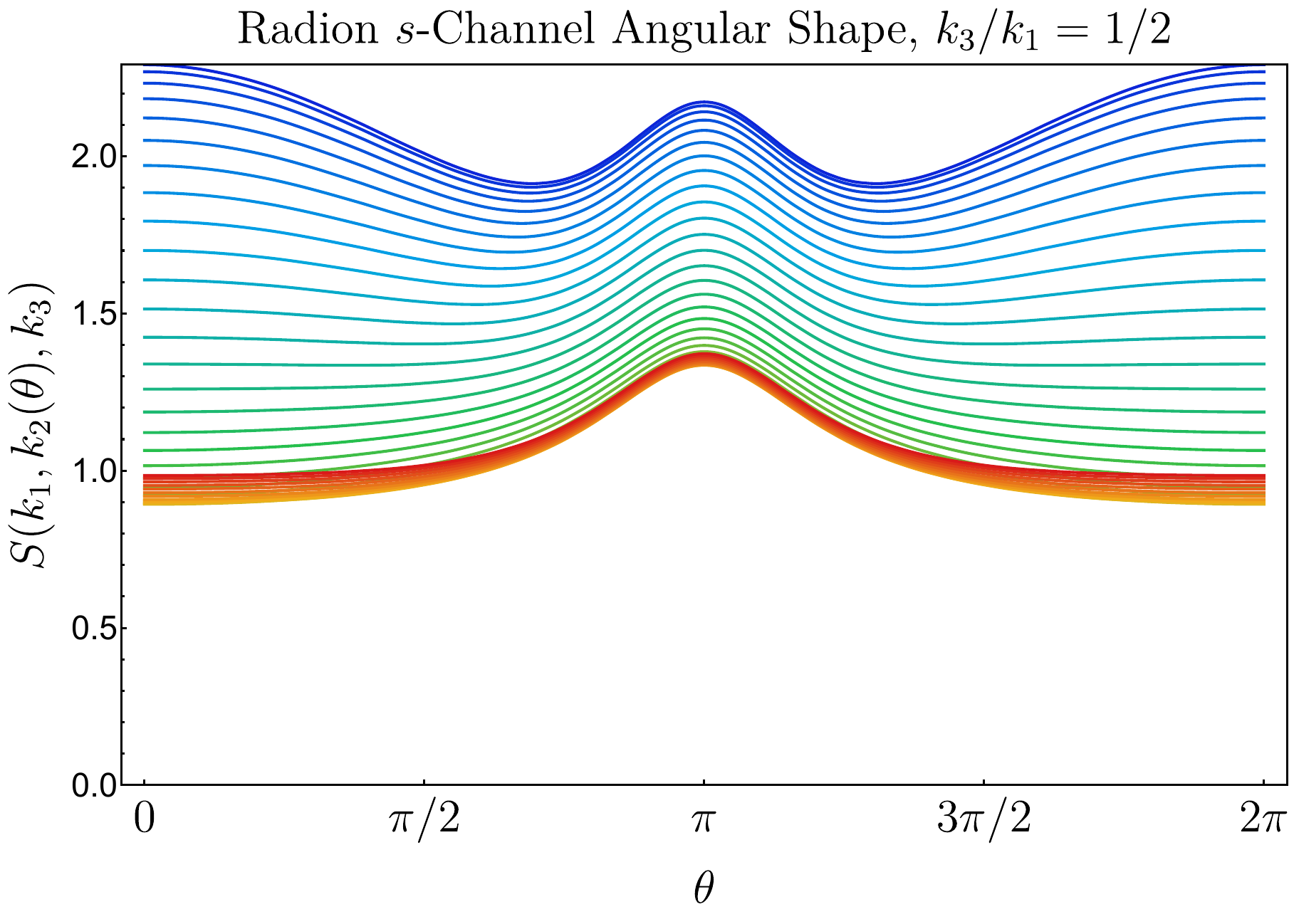}
}

\vspace{-0.6em}

\subfloat[All channels, $k_3/k_1=0.95$.\label{fig:radion-all-095}]{
  \includegraphics[width=0.47\textwidth]{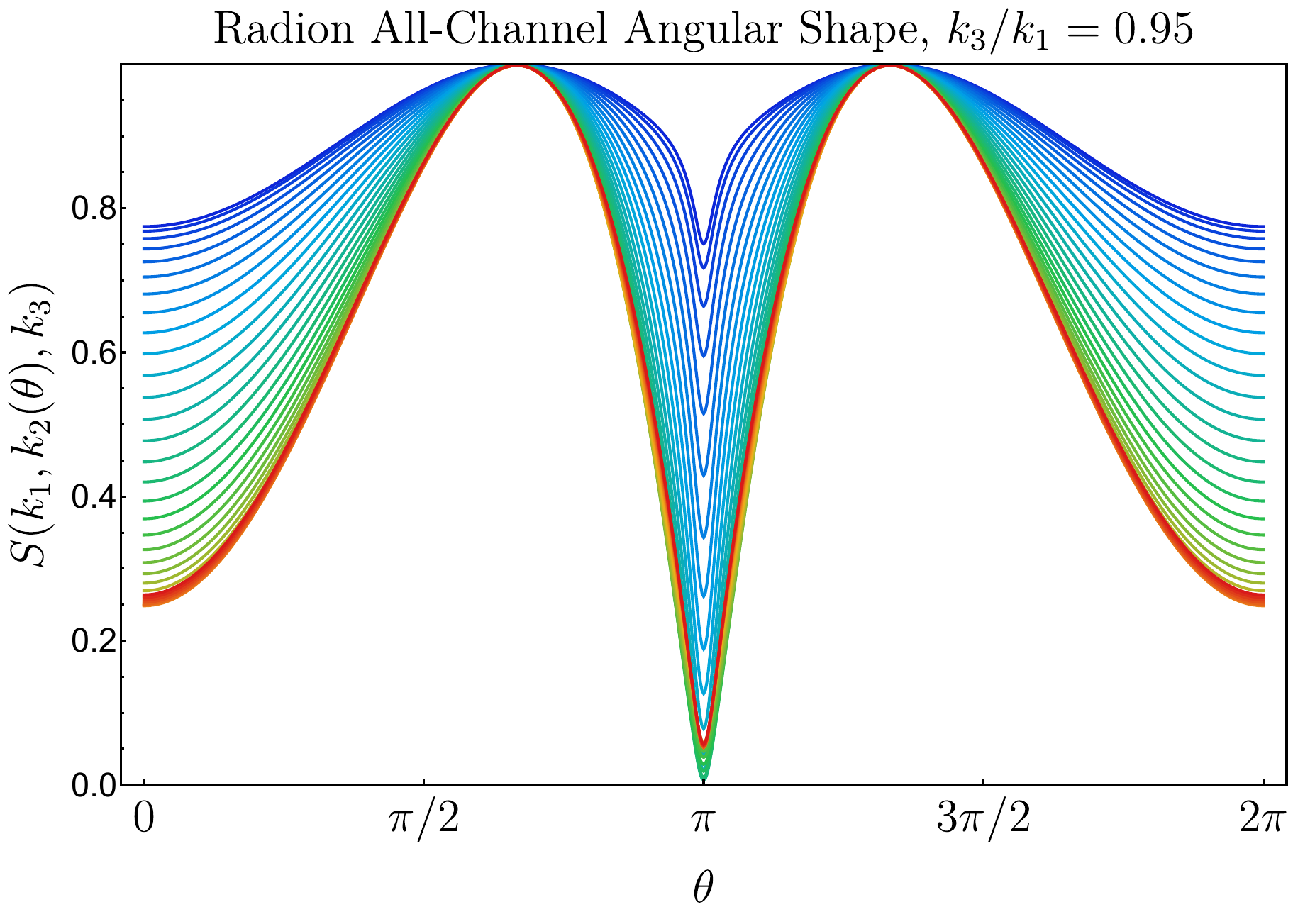}
}
\hfill
\subfloat[$s$-channel, $k_3/k_1=0.95$.\label{fig:radion-s-095}]{
  \includegraphics[width=0.47\textwidth]{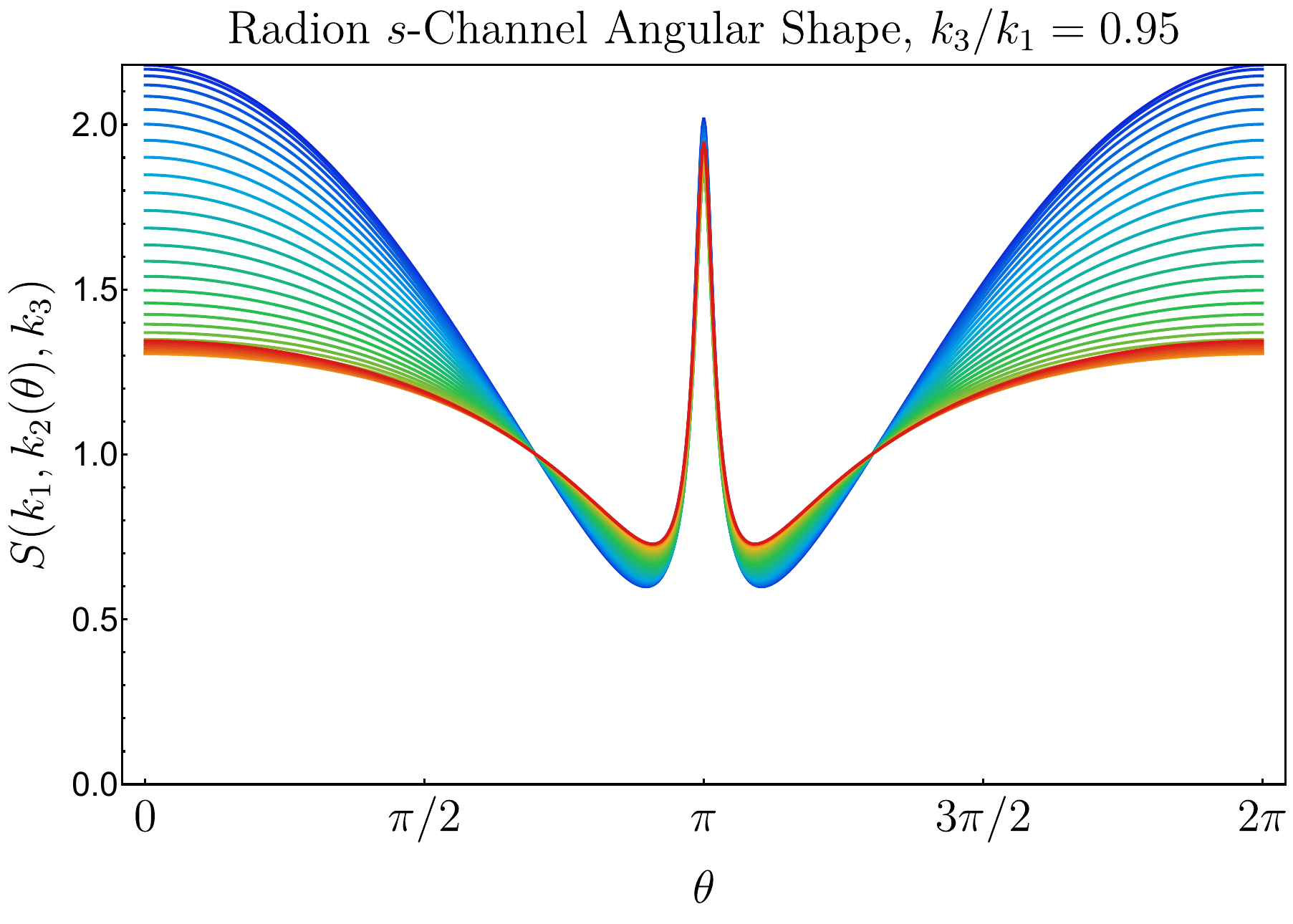}
}

\vspace{-0.8em}

\caption{Angular dependence of the radion shape function for the all-channel sum and the $s$-channel. Each curve corresponds to a mass in the range $\mu\in[0.1,3.5]$, increasing in steps of $0.1$, from blue to red. None of these crosses $S=0$.}
\label{fig:radion-angular-mass-scan}
\end{figure}
As mentioned at the end of the caption of Fig.~\ref{fig: KK graviton angular dependence comparison}, the radion shape function never crosses $S = 0$. A numerical evidence is presented in Fig.~\ref{fig:radion-angular-mass-scan}. Firstly, notice the six plots that display different ratios of $k_3/k_1$ for the same angular parameterization of the radion shape function as above. The graphs are a scan over the mass range $\mu\in [0.1,3.5]$ in steps of $0.1$, with colors changing from blue at low mass to red at high mass. 
None of the mass values we have been considering for the radion cross $S = 0$, which is a non-trivial difference compared to the shape function generated by massive spinning particles such as KK-graviton.

\begin{figure}[H]
\centering
\subfloat[Radion shape function.\label{fig:sub1squeeze}]{
  \includegraphics[width=0.47\columnwidth]{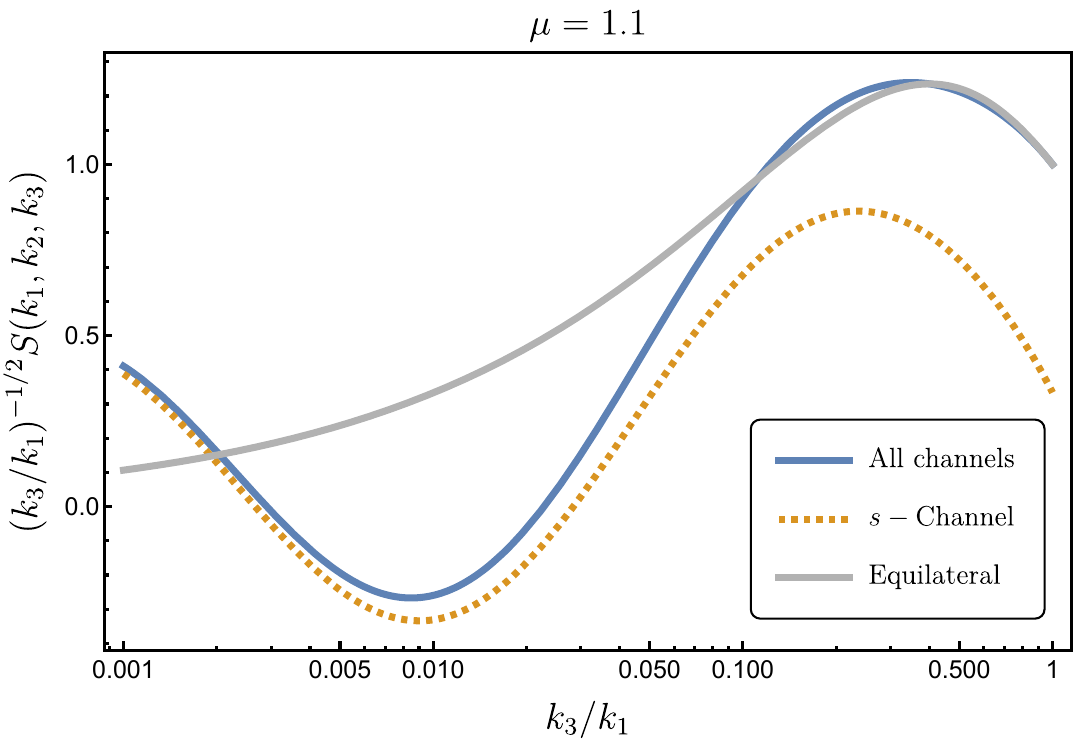}
}
\hfill
\subfloat[KK graviton shape function.\label{fig:sub2squeeze}]{
  \includegraphics[width=0.47\columnwidth]{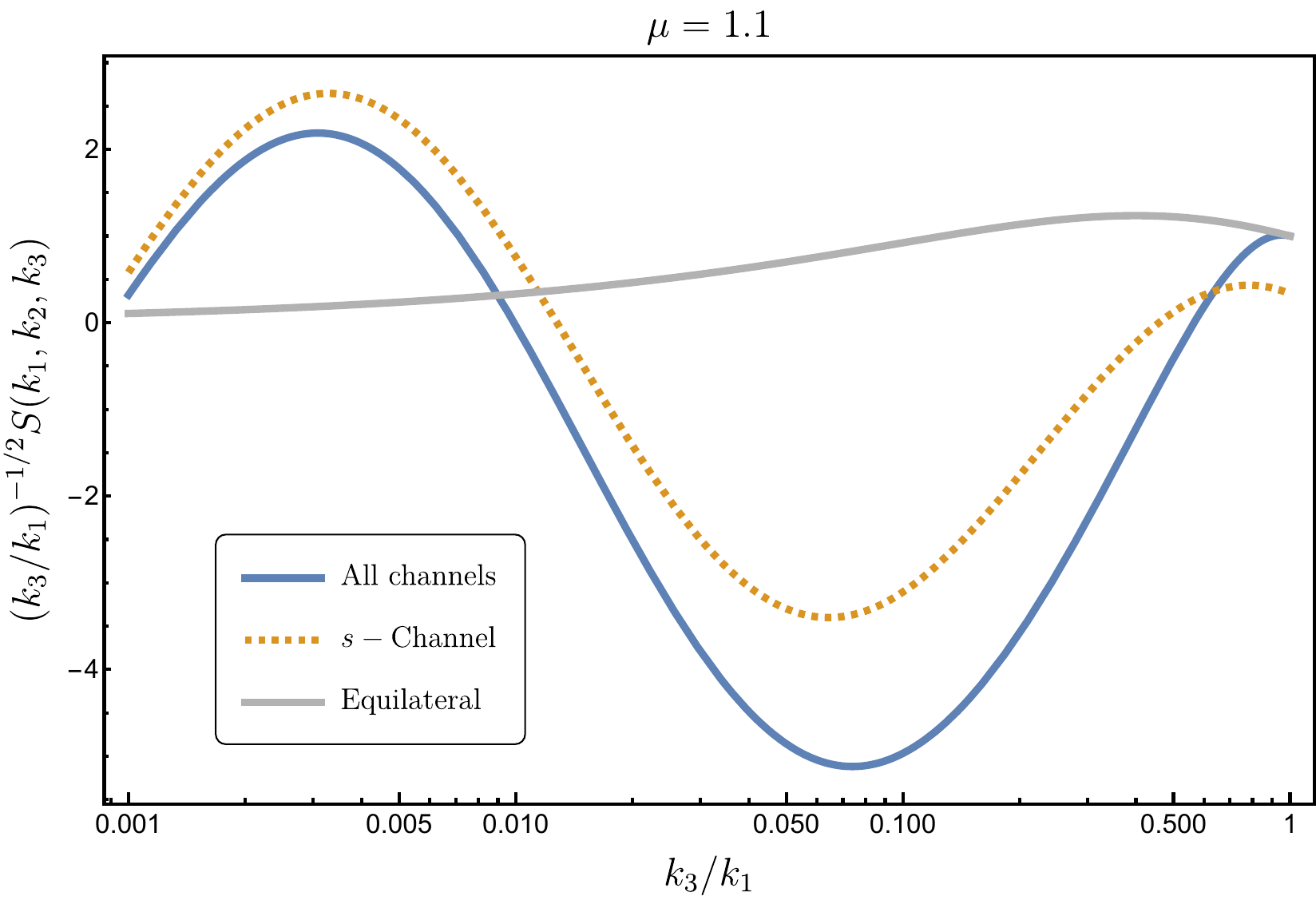}
}
\caption{Various shape functions. The oscillation (or $s$-channel) shape function and sum of all channels have been normalized with respect to the all channels shape function, while the equilateral shape has been normalized to itself.}
\label{fig:KK graviton and radion squeezed limit comparison}
\end{figure}

In the squeezed limit, the all-channel contribution gives a non-oscillatory component of the total shape.
We can also compare with the usual equilateral configuration that is produced by a massless scalar operator of the form $(\dot{\delta\sigma})^3$ producing a contact diagram and the following shape function \cite{Kumar:2026dih}
\begin{equation}
    S_\text{equil}(k_1,k_2,k_3) = 27\frac{k_1k_2k_3}{(k_1+k_2+k_3)^3}\label{eq: equilateral shape function}
\end{equation}
where the factor of $27$ comes from normalizing to the equilateral configuration. The comparison for the radion and the KK graviton can be seen in Fig.~\ref{fig:KK graviton and radion squeezed limit comparison}.

\section{Massive Spin-2 in the EFT of Inflation}\label{app: EFTofInf argument for spin-s}
In this section we discuss how the interactions in Eqs.~\eqref{eq: quadratic mixing int} and~\eqref{eq: cubic mixing} arise in the Goldstone EFT of inflation description~\cite{Cheung:2007st}.
We start from the term 
\es{}{
S \supset \int \D^4 x\sqrt{-g} f(t) \tilde{h}^{00}
}
in the unitary gauge.
Such a term is invariant under spatial diffeomorphism and hence is allowed.
In this unitary gauge, there are scalar perturbations of the metric, and it is more convenient to work directly with the Goldstone boson $\pi$ of spontaneously broken time translation invariance.
This can be introduced by doing the Stückelberg trick via a diffeomorphism, $t\rightarrow t + \pi(t,\vec{x})$ and $\vec{x}\rightarrow \vec{x}$.
Under this transformation,
\es{}{
\tilde{h}^{00} \rightarrow (\delta_\mu^0 + \partial_\mu\pi)(\delta_\nu^0 + \partial_\nu\pi)\tilde{h}^{\mu\nu} = \tilde{h}^{00} + 2 \partial_\mu\pi \tilde{h}^{\mu 0}  + \partial_\mu\pi \partial_\nu\pi \tilde{h}^{\mu\nu}.
}
As discussed in~\cite{Cheung:2007st}, for energies $E^2\gg |\dot{H}|$, we can ignore the mixing of $\pi$ with the metric fluctuations (the decoupling limit).
This means in the $\pi$-language, we can use the unperturbed metric $\bar{g}^{\mu\nu}$ to perform various contractions.
We also assume $f(t)$ is approximately a constant and hence drop its derivatives.
The action in the $\pi$-language is then given by,
\es{eq:pi-gauge}{
S \supset \int \D^4 x\sqrt{-\bar{g}} f (\tilde{h}^{00} + 2 \partial_\mu\pi \tilde{h}^{\mu 0}  + \partial_\mu\pi \partial_\nu\pi \tilde{h}^{\mu\nu}).
}
For the rest of this section, we assume the mass of the spin-2 field $\tilde{h}^{\mu\nu}$ is $m \gg H$, just to simplify the algebra. 

Working with the background metric $\D s^2 = -\D t^2 + a(t)^2 \D\vec{x}^2$, the first term in Eq.~\eqref{eq:pi-gauge} can be removed by performing the shift,
\es{}{
\tilde{h}^{ij} \rightarrow \tilde{h}^{ij} + B \delta^{ij}/a(t)^2,~~\tilde{h}^{00}\rightarrow \tilde{h}^{00} + A.
}
Under this shift, the tadpole generated from the Fierz-Pauli mass term for spin-2, cancels the tadpole of~\eqref{eq:pi-gauge} for $A=2B$ and $B = f/(3m^2)$. For $m\sim H$, there is an analogous transformation with slightly more complicated expressions for $A$ and $B$.
Thus after doing the above shift, there are no remaining tadpoles, and the relevant interactions come from
\es{eq:pi-gauge2}{
S \supset \int \D^4 x\sqrt{-\bar{g}} f (2 \partial_\mu\pi \tilde{h}^{\mu 0}  + \partial_\mu\pi \partial_\nu\pi \tilde{h}^{\mu\nu}) = \int \D^4 x\sqrt{-\bar{g}} f \left({2\over \eta} \partial_\mu\pi \tilde{h}^{\mu \eta}  + \partial_\mu\pi \partial_\nu\pi \tilde{h}^{\mu\nu}\right).
}
To get the second expression, we have passed from cosmic time $t$ to the conformal time $\eta$.
The cubic term in Eq.~\eqref{eq:pi-gauge2} directly gives the same interactions as in Eq.~\eqref{eq: cubic mixing}, with the identification $\pi \rightarrow \delta\sigma$.
To recover the quadratic mixing we write the second term as,
\es{}{
\int \D^4 x\sqrt{-\bar{g}} f {2\over \eta} \partial_\mu\pi \tilde{h}^{\mu \eta} = \int \D^4 x f {2\over \eta} \left(\partial_\eta\pi \tilde{h}_{\eta \eta} - \partial_i\pi\tilde{h}_{i \eta}\right).
}
Using the constraint equations for a massive spin-2 field,
\es{}{
\partial_\eta \tilde{h}_{\eta\eta} - \partial_i \tilde{h}_{i\eta} - {1\over \eta} \tilde{h}_{\eta\eta} - {1\over \eta} \tilde{h}_{ii} = 0,~~\tilde{h}_{\eta\eta} = \tilde{h}_{ii},
}
and doing integration-by-parts, we arrive at
\es{}{
\int \D^4 x f \left(-{2\over \eta^2}\right)\pi \tilde{h}_{\eta\eta}.
}
This has the same coupling structure as Eq.~\eqref{eq: quadratic mixing int}.

\section{Details of the 5D Setup}\label{sec:app5D}
\begin{figure}[H]
    \centering
    \includegraphics[width=0.47\linewidth]{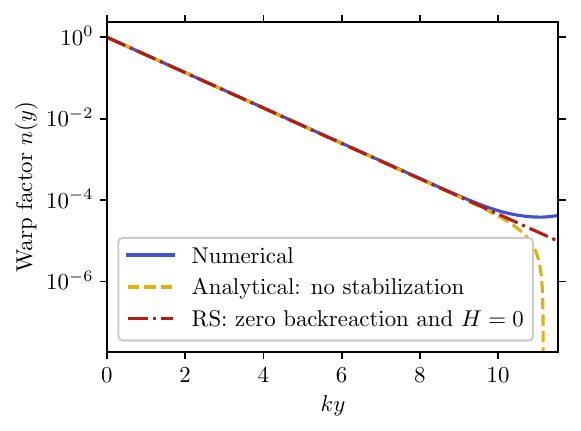}
    \includegraphics[width=0.47\linewidth]{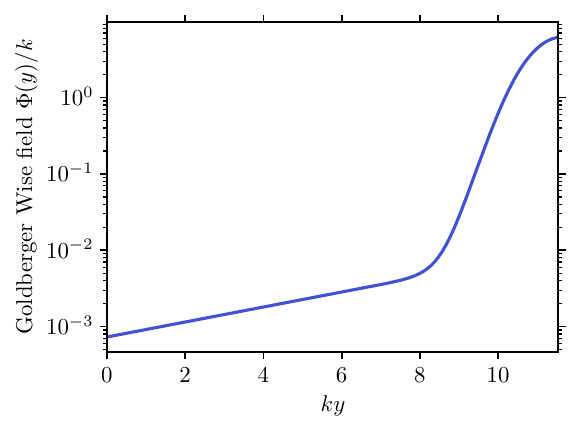}
    \caption{The numerical warp factor (left, solid blue) and the Goldberger Wise field profile (right), taking into account appropriate backreaction effects. For the left panel, we also show $n(y)$ in the absence of a stabilization (dashed yellow) and the standard RS result (dot-dashed red). For the various mass scales, we choose $M_5 = 1.2\times 10^{18}$~GeV, $k=6.4\times 10^{17}$~GeV, and $H=1.8\times 10^{13}$~GeV.
    For the dimensionless parameters, we choose $\epsilon=-0.85$, $\lambda_0=0.8$, $\lambda_L=0.12$, $f_0=5\times 10^{-4}$, $f_L=16$, $\tilde{V}=0$.}
    \label{fig:background_geom}
\end{figure}

The 5D action is given by 
\es{}{
S &= \int \D^4x \int_{-L}^{L} \D y \, \sqrt{-G} \bigg[(M_5^3 \mathcal{R}_5 - \Lambda_5 )+\bigg(-\frac{1}{2} G^{MN} \partial_M \Phi \, \partial_N \Phi- V(\Phi) - V_0(\Phi)\,\delta(y) - V_L(\Phi)\,\delta(y-L)\bigg)\bigg] + S_{\phi} + S_{\sigma}.
}
Here $V_0(\Phi)$ and $V_L(\Phi)$ are localized boundary terms for the Goldberger Wise (GW) field.
We will also use the convention that the potential energy density of the inflaton is absorbed into $V_0(\Phi)$, whereas that of $\sigma$ is negligible compared to $V_L(\Phi)$.
The Einstein equations are then
\es{eq:ee}{
6 M_5^3\left(H^2 - n'^2 - n n''\right) &=
n^2 \left(\Lambda_5 + \frac{1}{2}\Phi'^2 + V \right) +
n^2 \left(V_0 \delta(y) + V_L \delta(y-L)\right),\\
12 M_5^3\left(n'^2 - H^2\right) &=
n^2 \left(-\Lambda_5 + \frac{1}{2}\Phi'^2 - V\right),
}
where $n' = \D n/\D y$ and $\Phi' = \D \Phi/\D y$.
Integrating the first of Eq.~\eqref{eq:ee} across the two boundaries give the boundary conditions on the warp factor (for infinitesimally positive $\delta$)
\es{eq:nbc}{
6M_5^3\left[{n' \over n}\right]_{-\delta}^{+\delta} = - V_0,~~6 M_5^3\left[{n' \over n}\right]_{L+\delta}^{L-\delta} = V_L.
}
The equation of motion for the GW field is given by,
\es{eq:gw}{
\Phi'' + 4{n' \over n}\Phi'- {\D V\over \D \Phi} = 0.
}
We choose a free GW field for which $V(\Phi) = \epsilon k^2\Phi^2/2$.
Finally, we also need the boundary conditions on the GW field for which we split boundary potentials as
\es{}{
V_0 = 12 M_5^3 k + v_0(\Phi) + V_{\rm inf}(\phi),~~V_L = -12 M_5^3 k + v_L(\Phi) - \tilde{V},
}
where $v_0$ and $v_L$ are boundary localized potentials for the GW field, $V_{\rm inf}$ is the inflaton potential, and $\tilde{V}$ is any residual potential energy density on the IR boundary.
The boundary conditions on the GW field are thus given by,
\es{eq:gwbc}{
\left[\Phi'\right]_{-\delta}^{+\delta} = v_0'(\Phi),~~\left[\Phi'\right]_{L-\delta}^{L+\delta} = v_L'(\Phi).
}
We follow~\cite{Kumar:2025anx} for the specific choices of boundary potentials $v_0(\Phi)$ and $v_L(\Phi)$, and take
\es{}{
v_0(\Phi) = \lambda_0 k (k^{3/2}f_0-\Phi)^2 + \delta V_{\rm UV},~~v_L(\Phi) = \lambda_L k (k^{3/2}f_L-\Phi)^2.
}
Here $\lambda_{0,L}, f_{0,L}$ are dimensionless parameters, and $\delta V_{\rm UV}$ is a contribution that ensures $v_0(\Phi(y=0))=0$, so that the inflationary potential energy density solely comes from $V_{\rm inf}(\phi)$.
We solve a combination of Eq.~\eqref{eq:ee} and Eq.~\eqref{eq:gw}, along with boundary conditions~\eqref{eq:nbc} and~\eqref{eq:gwbc}, to determine the warp factor $n(y)$ and GW field profile $\Phi(y)$ simultaneously.
This automatically includes any backreaction due to the GW field, and improves the approximate semi-analytical treatment in Ref.~\cite{Kumar:2025anx}. 
We show the resulting warp factor and GW profile in Fig.~\ref{fig:background_geom}.
We also show the analytical warp factor in the absence of any stabilization, but non-zero $H$: $n(y) = \cosh(k y) - \sqrt{k^2 + H^2}\sinh(ky)/k$, which features a horizon where $n(y)=0$.
The standard RS result $n(y)=\exp(-ky)$ is also shown.
Our choices for the various dimensionful and dimensionless parameters are given in the caption of Fig.~\ref{fig:background_geom}.

\begin{figure}
    \centering
    \includegraphics[width=0.47\linewidth]{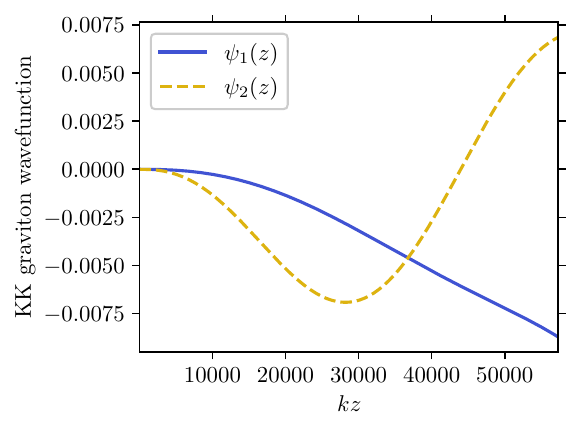}
    \includegraphics[width=0.47\linewidth]{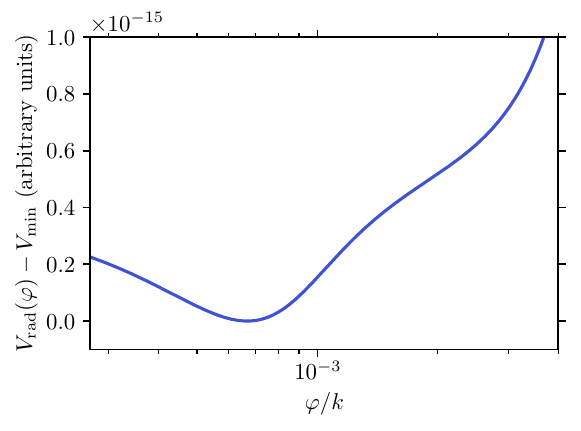}
    \caption{Wave function of the two lightest KK gravitons (left) and the radion potential near the minimum (right). We have subtracted a constant component of the radion potential to illustrate its shape. All the parameters are chosen to be the same as Fig.~\ref{fig:background_geom}.}
    \label{fig:kk_and_rad}
\end{figure}

With the background geometry determined, we use the parametrization in Eq.~\eqref{eq:5dmetric} to derive the properties of the radion and the KK graviton.
Upon a KK decomposition $h_{\mu\nu}(x,y) = \sum_\ell n^2(y) \chi_\ell(y)\tilde{h}_{\mu\nu,\ell}(x)$, the linearized Einstein equations give
\es{}{
(\square_{\rm dS}- m_\ell^2 - 2 H^2)\tilde{h}_{\mu\nu,\ell} &= 0,\\n^2{\D^2 \chi_\ell \over \D y^2} + 4 n {\D n \over \D y}{\D \chi_\ell \over \D y} + m_\ell^2 \chi_\ell &= 0,
}
where $\square_{\rm dS}$ is the Laplacian for 4D de Sitter (dS) space.
Thus, $\tilde{h}_{\mu\nu,\ell}$, for $\ell \geq 1$, indeed behaves as a massive spin-2 particle in dS.
We solve these equations, subject to Neumann boundary conditions $\D\chi_\ell/\D y = 0$ at $y=0,L$.
The zero mode $\chi_0$, for which $m_\ell = 0$, has a simple solution: $\chi_0={\rm constant}$.
This mode reproduces the standard 4D gravity and gives the relation $\mpl^2 = 2 M_5^3/k$ between the 4D Planck scale and the 5D scales.
To solve for $\chi_\ell$, it is convenient to do the transformation $\D z = \D y/n$ and $\psi_\ell(z)=n^{3/2}(y)\chi_\ell(y)$.
The wave functions $\psi_\ell(z)$ satisfies a Schrödinger-like equation and satisfy a normalization condition:
\es{}{
2 M_5^3 \int_{z_{\rm UV}}^{z_{\rm IR}} \D z \psi_{\ell}^*(z) \psi_{\ell'}(z) = {\mpl^2 \over 2}\delta_{\ell \ell'}.
}
The associated eigenvalues determine the KK graviton masses. 
We show the first two wave functions in Fig.~\ref{fig:kk_and_rad} (left), which correspond to masses $m_{1}\approx 1.57H$ and $m_{2}\approx 3.86 H$.

\begin{figure}
    \centering
    \includegraphics[width=0.75\linewidth]{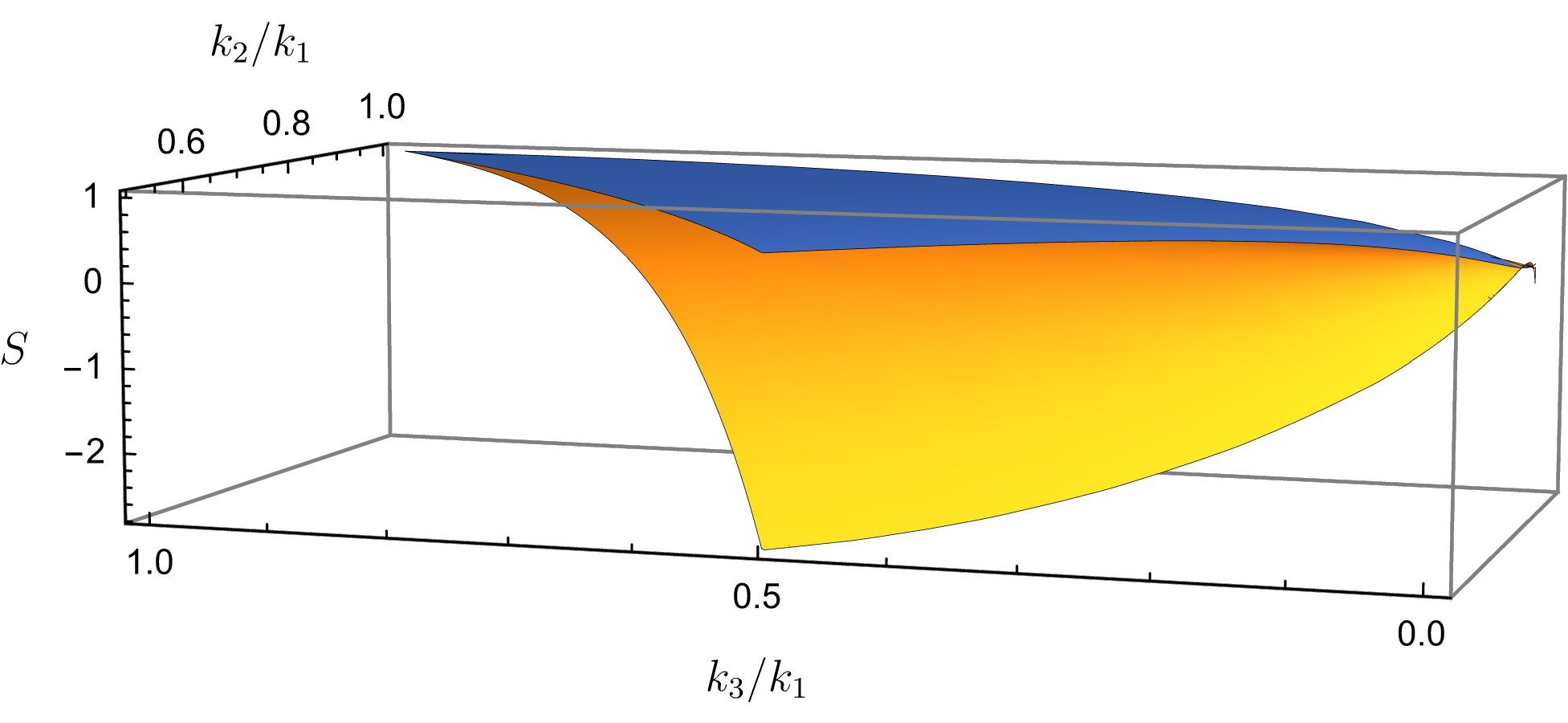}
    \caption{KK-graviton shape function in {\color{orange}{orange}} (excluding the contact term) and the radion shape function in {\color{blue}{blue}} at the benchmark mass values ($m_{\rm KK}\approx 1.57H$ and $m_{\rm rad}\approx 1.8H$). The shape functions have been normalized such that $S = 1$ at the equilateral point $k_1 = k_2 = k_3 =1$.}
    \label{fig:benchmark shape comparison}
\end{figure}

\begin{figure}
\centering
\subfloat[\label{fig:sub1p}]{
  \includegraphics[width=0.48\columnwidth]{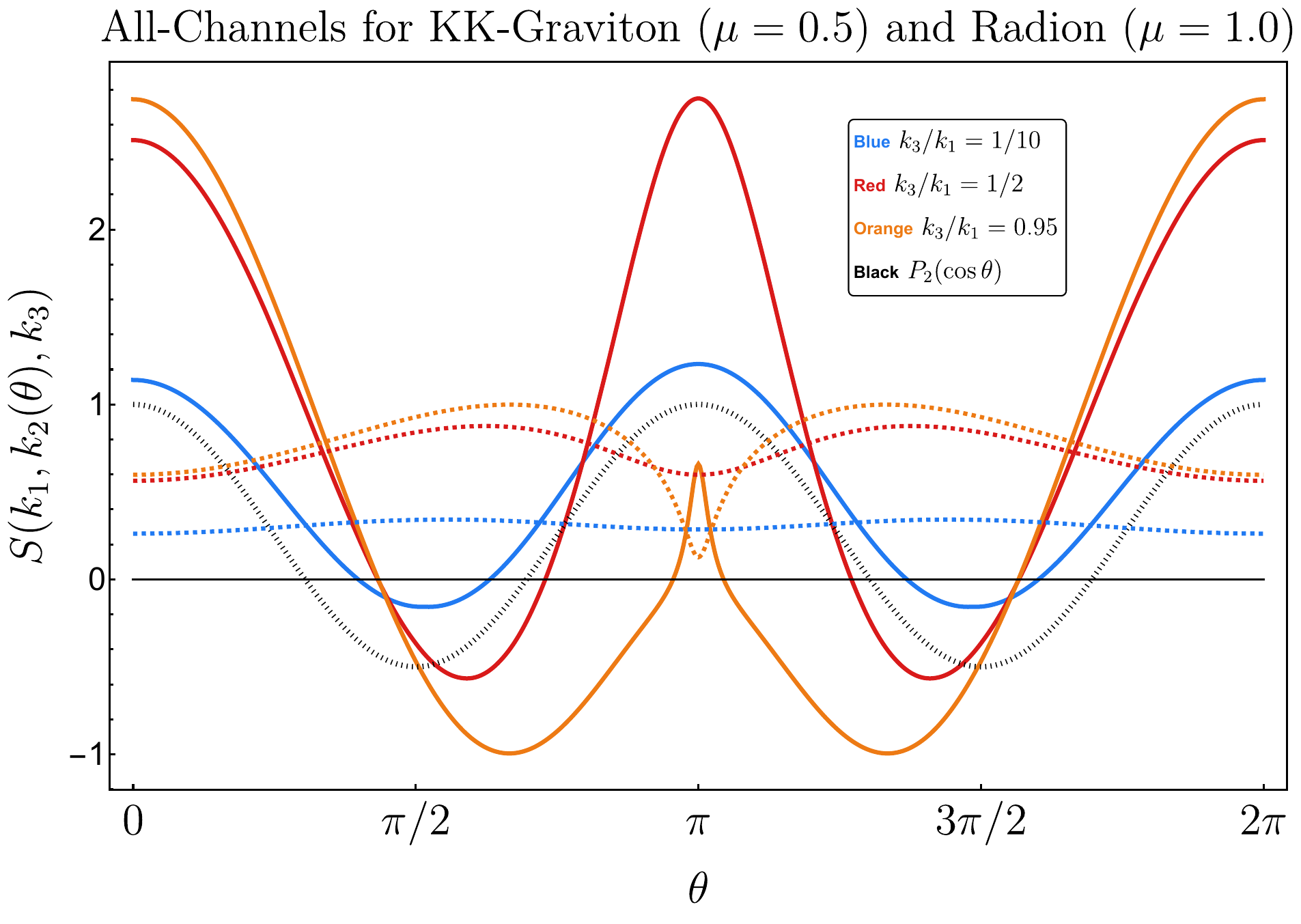}
}
\hfill
\subfloat[\label{fig:sub2p}]{
  \includegraphics[width=0.48\columnwidth]{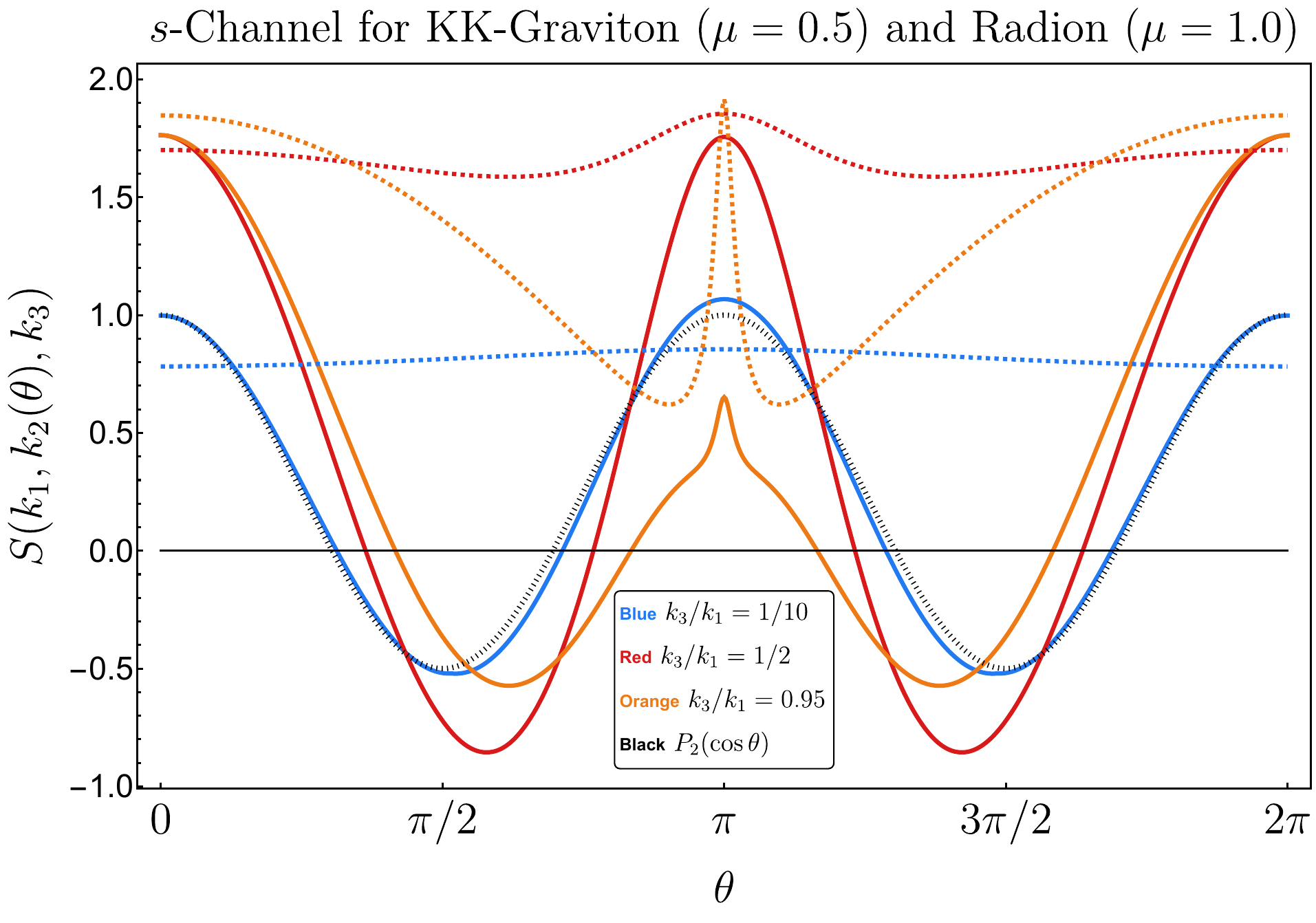}
}
\caption{Angular dependence of the shape functions for the KK-graviton and radion at the benchmark mass values. The KK-graviton shape function are the solid lines while the radion shape function are the dashed lines. Note that in \ref{fig:sub1p} the KK-graviton shape function has been scaled by $-1$ while in \ref{fig:sub2p} it has been scaled by $-1/2$.}
\end{figure}

In Ref.~\cite{Kumar:2025anx}, one of us derived the action for the canonically normalized radion field $\varphi = F_\varphi \exp(-k\pi r)$, with
\es{}{
F_{\varphi}^{2}= -\frac{24 M_5^3 e^{2kL}}{(kL)^2}
\int_{0}^{L} \D y y\, {\D n(y)\over \D y}\left(n + y  {\D n(y)\over \D y}\right).
}
In the absence of the GW stabilization, if the IR boundary is too close to the horizon $F_{\varphi}^{2}<0$, while including it ensures $F_{\varphi}^{2}>0$, along with a non-tachyonic radion.
The stabilizing effect of the GW field in preventing $n(y)$ going to zero can be seen in Fig.~\ref{fig:background_geom}.
The radion potential is given by~\cite{Kumar:2025anx},
\es{}{
\frac{V_{\rm rad}}{\sqrt{-\bar g}} = n^4(\pi r) \left[-12M_5^3 \frac{n'(\pi r)}{n(\pi r)} +v_L\!\bigl(\Phi(\pi r)\bigr) -12M_5^3 k -\tilde V \right] -12M_5^3 H^2 \int_0^{\pi r} \D y\, n(y)^2,
}
with $r = \log(F_\varphi/\varphi)/(k\pi)$.
For the parameter choice given in the caption of Fig.~\ref{fig:background_geom}, we find that the radion minimum occurs at $\langle\varphi\rangle/k\approx 6.7\times 10^{-4}$ (Fig.~\ref{fig:kk_and_rad} (right)), with $\langle r\rangle \approx 3.67/k$.
The radion mass is given by $m_{\rm rad}\approx 1.8 H$.
With the above ingredients, the quantities relevant for Eq.~\eqref{eq:int5D} can be obtained.
We find $n'(L)/(k n(L))\approx 0.43$, $\langle\varphi\rangle \approx 23H$, relevant for the radion couplings.
While for the lightest KK graviton, we find $\chi_1/\mpl \approx -0.24/H$.
We show the shape functions for these values of the radion and the lightest KK graviton mass in Figs.~\ref{fig:benchmark shape comparison},~\ref{fig:sub1p}, and~\ref{fig:sub2p}.

\section{Density Perturbations from Modulated Reheating}\label{app:boltzmann}
\begin{figure}
    \centering
    \includegraphics[width=0.5\linewidth]{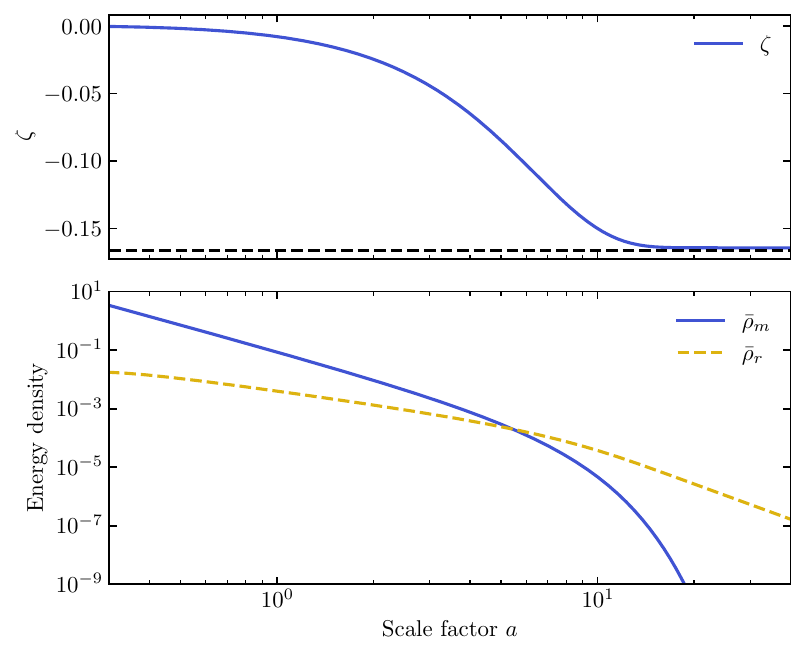}
    \caption{Generating curvature perturbation from modulating reheating. We choose $\delta\Gamma/\Gamma=1$.}
    \label{fig:mod}
\end{figure}

To understand how modulated reheating sources density perturbations, we model the end of inflation as the inflaton diluting as a matter field and then decaying into radiation~\cite{Dvali:2003em}.
This is governed by the equations,
\es{}{
3\mathcal{H}^2-a^2 
\left(\bar{\rho}_m + \bar{\rho}_r\right)/\mpl^2
&= 0,\\
\bar{\rho}_r'+ 4\mathcal{H}\bar{\rho}_r-a\Gamma\,\bar{\rho}_m
&= 0,\\
\bar{\rho}_m' + 3\mathcal{H}\bar{\rho}_m +a\Gamma\,\bar{\rho}_m
&=0\,.
}
Here $\bar{\rho}_{m}$ and $\bar{\rho}_{r}$ are homogeneous energy densities in the inflaton (matter) and its decay products (radiation), respectively, with ${\cal H}$ the conformal Hubble parameter and $\Gamma$ the homogeneous inflaton decay rate.
In this section, we use a prime to denote derivatives with respect to the conformal time.
The evolution of the homogeneous energy densities are shown in Fig.~\ref{fig:mod}.
To understand how a spatial variation in $\Gamma$ sources curvature perturbation, we work in the conformal Newtonian gauge in which the scalar metric perturbations, in the absence of anisotropic stress, are given as,
\es{}{
\D s^2 = a^2(\eta) (-(1+2\Phi)\D \eta^2  + (1-2\Phi)\D \vec{x}^2).
}
The gauge invariant curvature perturbation on uniform density hypersurfaces, $\zeta = -\Phi-{\cal H} \delta\rho/{\bar{\rho}}'$, is determined in terms of the total density fluctuation $\delta\rho$ and its homogeneous value $\bar{\rho}$, where all these quantities are evaluated after inflation.
To determine $\zeta$, we solve the Boltzmann equations involving the matter density ($\delta_m = \delta\rho_m/\bar{\rho}_m$), matter velocity ($\theta_m$), radiation density ($\delta_r$), and radiation velocity ($\theta_r$), and $\Phi$.
These equations are given by,
\es{}{
\delta_r' -4\Phi' +\frac{4}{3}\theta_r -a\Gamma \frac{\bar{\rho}_m}{\bar{\rho}_r} \left(\Phi +\delta_m -\delta_r \right) -\frac{\bar{\rho}_m}{\bar{\rho}_r}\,a\,\delta\Gamma &= 0, \, \\
\delta_m'-3\Phi' +\theta_m +a\Gamma \Phi +a\,\delta\Gamma &= 0 \,,\\
3\mathcal{H} \left(\Phi' + \mathcal{H}\Phi\right) +\frac{a^2}{2\mpl^2}\,  \left(\bar{\rho}_m \delta_m + \bar{\rho}_r \delta_r\right) + k^2\Phi &= 0.
}
Since we are interested in the superhorizon limit ($k\ll {\cal H}$), where the fluctuations in the decay rate $\delta\Gamma$ source $\zeta$, we set $k=0$ along with $\theta_m=\theta_r=0$.
We solve these equations with initial conditions $\delta_m=\delta_r=\Phi=0$ and $\delta\Gamma/\Gamma = 1$.
The resulting $\zeta$ is shown in Fig.~\ref{fig:mod}.
With our choice of $\Gamma \ll H$ initially, we find $\zeta_{\rm final}\approx -1/6$ (dashed black line in Fig.~\ref{fig:mod}), consistent with Ref.~\cite{Dvali:2003em}.

\end{document}